\documentclass[12pt]{article}

\usepackage{cite}

\author{Yu.~M.~Zinoviev
       \thanks{E-mail address: Yurii.Zinoviev@ihep.ru} \\[0.5cm]
        {\it Institute for High Energy Physics} \\
        {\it of National Research Center "Kurchatov Institute"} \\
        {\it Protvino, Moscow Region, 142280, Russia}}

\title{Partially massless spin 5/2 and supersymmetry}

\date{}

\begin{document}

\maketitle

\begin{abstract}
We elaborate on the partially massless spin 5/2 supermultiplet, which
contains partially massless spin 5/2, massless and partially massless
spin 2, as well as massless spin 3/2. We consider the global
supertransformations connecting partially massless spin 5/2 to its two
possible superpartners, massless and partially massless spin 2, and
make them local by switching the interaction with the massless
gravitino. We use a frame-like gauge-invariant formalism to describe
free fields and the Fradkin-Vasiliev formalism to construct
interactions, Due to the presence of the Stueckelberg fields in the
gauge-invariant description of massive and partially massless fields,
we face ambiguities related to field redefinitions. We use this
freedom to simplify calculations. At the same time, we demonstrate how
these ambiguities can be resolved using unfolded equations.
\end{abstract}

\thispagestyle{empty}
\newpage
\setcounter{page}{1}

\section{Introduction}

Partially massless fields \cite{DW01,DW01a,DW01c,Zin01,Met06,Met22a}
correspond to rather exotic representations of de Sitter algebra that
have no direct analogies among the representations of Poincare
algebra\footnote{In a flat limit, they become reducible and decompose
into a number of massless representations.}. Till now most papers
have been devoted to the partially massless spin 2 and the possibility
of constructing partially massless gravity or bigravity. The cubic
vertices describing self-interaction of these fields and their
interaction with massless graviton exist \cite{Zin06,Zin14}, but all
attempts to go beyond the cubic level have not been successful and
mostly produced some no-go statements 
\cite{RR12,HSS12,HSS12d,DSW13,RHRT13,JLT14,JR15,GHJMR16,BDGT19,JMP19,ST21,BGPT24}.
There is an interesting connection between partially massless spin 2
and conformal gravity \cite{DJW12,DJW13,HSS13} (see e.g. \cite{Met14}
and references therein for generalization to arbitrary conformal
fields). One more interesting subject is the possibility to extend the
Vasiliev theory to include both massless and partially massless
fields. Naturally, the first step is a search for an appropriate
infinite-dimensional algebra and there are some candidates
\cite{BG13,JM15,BD24}. 

The partially massless fields in anti de Sitter space are non-unitary,
but it has been shown that supermultiplets with these fields do
exist in $AdS_4$ \cite{GHR18,BKhSZ19a,BHKP21,BGHR21}. There
is an interesting connection with the superconformal gravity and
higher spins (see e.g. \cite{FKL18,KP19,BHKP21} and references
therein). Among these supermultiplets there is an interesting subclass
that contains both massless and partially massless fields.
The first example is the partially massless spin 2 supermultiplet
which was investigated recently in  \cite{Zin18a,BLT23,Zin24b,BLT24}.
In this work, we deal with a partially massless spin 5/2
supermultiplet that contains partially massless spin 5/2, massless and
partially massless spin 2 and massless spin 3/2. We consider global
supertransformations that connect partially massless spin 5/2 to its
two possible superpartners (massless or partially massless spin 2) and
make these global supersymmetries local by introducing interactions
with massless spin 3/2 gravitino.

A frame-like formalism for the partially massless bosons has been
proposed by Skvortsov and Vasiliev in \cite{SV06}. Note that in this
formalism, the case with depth $t=0$ (which includes partially
massless spin 2) is a special case. In this situation, it is
impossible to construct a first order Lagrangian that could be called
a frame-like one, while the only possible second order Lagrangian is
equivalent to a metric-like formalism. A general frame-like formalism
for massive higher spins (including all possible partially massless
limits) was developed in \cite{Zin08b,PV10} and then adapted to the
multispinor formalism in \cite{KhZ19}. In \cite{KhZ19}, it was
demonstrated that Skvortsov-Vasiliev formalism is simply a gauge fixed
version of the general one and using this knowledge, their formalism
was extended to the fermions as well. Note that, in general, the two
procedures, partial gauge fixing and interactions switching, do not
commute. One of goals of this paper is to compare the results obtained
using both possible descriptions of partially massless spin 5/2.

One of the general features of the frame-like formalism is the
appearance of so-called extra fields, which do not enter into the free
Lagrangian but are necessary for constructing a complete set of
gauge invariant objects. As a result, the usual constructive approach
to interactions cannot be used, and the most efficient method is the
so-called Fradkin-Vasiliev formalism \cite{FV87,FV87a,Vas11,Zin24a}.
Note also that the gauge invariant description of massive and
partially massless fields includes Stueckelberg zero-forms, which
leads to ambiguities associated with field redefinitions
\cite{BDGT18,Zin24a}. In the Fradkin-Vasiliev formalism these
redefinitions produce combinations of abelian and non-abelian vertices
that vanish on-shell (up to total derivatives) and do not alter the
on-shell part of the cubic vertex. Therefore, they must be taken into
account in order to accurately estimate the number of independent
vertices. Regarding supersymmetry, these ambiguities can be resolved
using unfolded equations, as demonstrated for massive supermultiplets
in \cite{KhZ20,Zin24}.

The paper is organized as follows. In Section two, we provide all
necessary kinematics for partially massless and massless spin 5/2,
partially massless and massless spin 2 and massless spin 3/2.
In Section three, we consider the massless supermultiplet $(5/2,2)$ as
a simple illustration of the general approach. In Section four, we
consider massless spin 2 as a superpartner for partially massless spin
5/2 using two different descriptions for later. We show that at
least in this particular case they produce consistent results.
Section five is dedicated to the case of partially massless spin 2 as
a superpartner. Here we restrict ourselves to the Skvortsov-Vasiliev
formalism. Finally, in the Appendix, we show how ambiguity in
supertransformations related to field redefinitions can be resolved
using unfolded equations.

\section{Kinematics}

In this Section, we provide all necessary kinematical information on
the frame-like multispinor formalism \cite{KhZ19} we use. In this
formalism all objects are forms (fields are one-forms or zero-forms,
gauge invariant curvatures are two-forms or one-forms, while all the
terms in the Lagrangians are four-forms) with some number of
symmetric dotted and undotted spinor indexes 
$\alpha,\dot\alpha = 1,2$. The coordinate-free description of (anti)
de Sitter space is obtained using the background frame
$e^{\alpha\dot\alpha}$ and the covariant external derivative $D$
(which contains a background Lorentz connection) that satisfy
\begin{equation}
D \wedge e^{\alpha\dot\alpha} = 0, \qquad
D \wedge D \zeta^\alpha = 2\Lambda E^\alpha{}_\beta \zeta^\beta.
\end{equation}
Here basic two-forms $E^{\alpha(2)}$, $E^{\dot\alpha(2)}$ are defined
as follows
\begin{equation}
e^{\alpha\dot\alpha} \wedge e^{\beta\dot\beta} =
\epsilon^{\alpha\beta} E^{\dot\alpha\dot\beta}
+ e^{\dot\alpha\dot\beta} E^{\alpha\beta}.
\end{equation}
In what follows all wedge product signs are omitted.

\subsection{Partially massless spin 5/2}

Partially massless spin 5/2 has four physical helicities 
$(\pm \frac{5}{2}, \pm \frac{3}{2})$ and to describe it we introduce 
one-forms $\Phi^{\alpha(2)\dot\alpha} + h.c.$ and 
$\Phi^\alpha + h.c.$. The free Lagrangian has the form
\begin{eqnarray}
{\cal L}_0 &=& \Phi_{\alpha\beta\dot\alpha} e^\beta{}_{\dot\beta}
D \Phi^{\alpha\dot\alpha\dot\beta} - \Phi_\alpha 
e^\alpha{}_{\dot\alpha} D \Phi^{\dot\alpha} + a_0 
\Phi_{\alpha(2)\dot\alpha} E^{\alpha(2)} \Phi^{\dot\alpha} \nonumber 
\\
 && + \frac{M}{4} (3 \Phi_{\alpha\beta\dot\alpha} E^\beta{}_\gamma
\Phi^{\alpha\gamma\dot\alpha} - \Phi_{\alpha(2)\dot\alpha}
E^{\dot\alpha}{}_{\dot\beta} \Phi^{\alpha(2)\dot\beta}) 
- \frac{3M}{2} \Phi_\alpha E^\alpha{}_\beta \Phi^\beta + h.c.,
\end{eqnarray}
where
$$
a_0{}^2 = - \frac{15}{4}\lambda^2, \qquad M^2 = \lambda^2.
$$
This Lagrangian is invariant under the following gauge
transformations:
\begin{eqnarray}
\delta \Phi^{\alpha(2)\dot\alpha} &=& D \rho^{\alpha(2)\dot\alpha}
+ e_\beta{}^{\dot\alpha} \rho^{\alpha(2)\beta} + 
\frac{M}{2} e^\alpha{}_{\dot\beta} \rho^{\alpha\dot\alpha\dot\beta}
 + \frac{a_0}{3}e^{\alpha\dot\alpha} \rho^\alpha, \nonumber \\
\delta \Phi^\alpha &=& D \rho^\alpha + a_0 e_{\beta\dot\alpha}
\rho^{\alpha\beta\dot\alpha} + \frac{3M}{2} e^\alpha{}_{\dot\alpha}
\rho^{\dot\alpha}. 
\end{eqnarray}
Note that beyond the usual gauge transformations involving the
parameters $\rho^{\alpha(2)\dot\alpha}$ and $\rho^\alpha$, we have
algebraic transformations with the parameter $\rho^{\alpha(3)}$, and
this is important for having a correct number of physical degrees of
freedom. As a result, in order to construct a complete set of gauge
invariant objects, we need extra fields: a one-form 
$\Phi^{\alpha(3)} + h.c.$ and a zero-form $\phi^{\alpha(3)} + h.c.$
with the gauge transformations
\begin{equation}
\delta \Phi^{\alpha(3)} = D \rho^{\alpha(3)}, \qquad
\delta \phi^{\alpha(3)} = \rho^{\alpha(3)}.
\end{equation}
Then we obtain:
\begin{eqnarray}
{\cal F}^{\alpha(3)} &=& D \Phi^{\alpha(3)} + 2\lambda^2 
E^\alpha{}_\beta \phi^{\alpha(2)\beta}, \nonumber \\
{\cal F}^{\alpha(2)\dot\alpha} &=& D \Phi^{\alpha(2)\dot\alpha} 
+ e_\beta{}^{\dot\alpha} \Phi^{\alpha(2)\beta} + \frac{M}{2}
e^\alpha{}_{\dot\beta} \Phi^{\alpha\dot\alpha\dot\beta}
 + \frac{a_0}{3} e^{\alpha\dot\alpha} \Phi^\alpha, \nonumber \\
{\cal F}^\alpha &=& D \Phi^\alpha + a_0 e_{\beta\dot\alpha}
\Phi^{\alpha\beta\dot\alpha} + \frac{3M}{2}
e^\alpha{}_{\dot\alpha}\Phi^{\dot\alpha} - 2a_0 E_{\beta(2)}
\phi^{\alpha\beta(2)}, \\
{\cal C}^{\alpha(3)} &=& D \phi^{\alpha(3)} - \Phi^{\alpha(3)}.
\nonumber
\end{eqnarray}
Note that on-shell we have
\begin{equation}
{\cal F}^{\alpha(2)\dot\alpha} \approx 0, \qquad
{\cal F}^\alpha \approx 0.
\end{equation}
As a result the remaining curvatures satisfy differential identities
\begin{equation}
D {\cal F}^{\alpha(3)} \approx 2\lambda^2 E^\alpha{}_\beta
{\cal C}^{\alpha(2)\beta}, \qquad D {\cal C}^{\alpha(3)} \approx
- {\cal F}^{\alpha(3)},
\end{equation}
as well as algebraic ones
\begin{equation}
e_\beta{}^{\dot\alpha} {\cal F}^{\alpha(2)\beta} \approx 0, \qquad
E_{\beta(2)} {\cal C}^{\alpha\beta(2)} \approx 0.
\end{equation}
Using these gauge invariant curvatures we can rewrite the free
Lagrangian in the explicitly gauge invariant form:
\begin{eqnarray}
{\cal L}_0 &=& h_1 {\cal F}_{\alpha(3)} {\cal F}^{\alpha(3)} - 
\frac{1}{2M} {\cal F}_{\alpha(2)\dot\alpha} 
{\cal F}^{\alpha(2)\dot\alpha} + \frac{1}{3M} {\cal F}_\alpha 
{\cal F}^\alpha \nonumber \\
 && - \frac{1}{M} {\cal C}_{\alpha(2)\beta} e^\beta{}_{\dot\alpha}
{\cal F}^{\alpha(2)\dot\alpha} + h_2{\cal C}_{\alpha(2)\beta}
E^\beta{}_\gamma {\cal C}^{\alpha(2)\gamma} + h.c.,
\end{eqnarray}
where
$$
6\lambda^2 h_1 - h_2 = \frac{1}{M}.
$$
Let us turn to the unfolded equations \cite{KhZ19}. As it has been
already noted for the physical fields we have
\begin{eqnarray}
0 &=& D \Phi^{\alpha(2)\dot\alpha} + e_\beta{}^{\dot\alpha}
\Phi^{\alpha(2)\beta} + \frac{M}{2} e^\alpha{}_{\dot\beta}
\Phi^{\alpha\dot\alpha\dot\beta} + \frac{a_0}{3} e^{\alpha\dot\alpha}
\Phi^\alpha, \nonumber \\
0 &=& D \Phi^\alpha + a_0 e_{\beta\dot\alpha}
\Phi^{\alpha\beta\dot\alpha} + \frac{3M}{2} e^\alpha{}_{\dot\alpha}
\Phi^{\dot\alpha} - 2a_0 E_{\beta(2)} \phi^{\alpha\beta(2)},
\end{eqnarray}
while for the extra fields we have
\begin{eqnarray}
0 &=& D \Phi^{\alpha(3)} + 2\lambda^2 E^\alpha{}_\beta
\phi^{\alpha(2)\beta} + E_{\beta(2)} Y^{\alpha(3)\beta(2)}, \nonumber 
\\
0 &=& D \phi^{\alpha(3)} - \Phi^{\alpha(3)} + e_{\beta\dot\alpha}
\phi^{\alpha(3)\beta\dot\alpha}.
\end{eqnarray}
Here $Y^{\alpha(5)}$ and $\phi^{\alpha(4)\dot\alpha}$ are the first
two representatives of the two infinite chains of the gauge invariant
zero-forms with the equations
\begin{eqnarray}
0 &=& D Y^{\alpha(5+k)\dot\alpha(k)} + e_{\beta\dot\beta}
Y^{\alpha(5+k)\beta\dot\alpha(k)\dot\beta} + c_{1,k}
e^\alpha{}_{\dot\beta} \phi^{\alpha(4+k)\dot\alpha(k)\dot\beta}
+ d_{1,k} e^{\alpha\dot\alpha} Y^{\alpha(4+k)\dot\alpha(k-1)},
\nonumber \\
0 &=& D \phi^{\alpha(4+k)\dot\alpha(k+1)} + e_{\beta\dot\beta}
\phi^{\alpha(4+k)\beta\dot\alpha(k+1)\dot\beta} + c_{2,k}
e_\beta{}^{\dot\alpha} Y^{\alpha(4+k)\beta\dot\alpha(k)} + d_{2,k}
e^{\alpha\dot\alpha} \phi^{\alpha(3+k)\dot\alpha(k)}, 
\end{eqnarray}
where
\begin{eqnarray}
c_{1,k} &=& \frac{15\lambda^2}{(k+5)(k+6)}, \qquad
d_{1,k} = \frac{(k+2)(k+4)}{(k+1)(k+5)}\lambda^2, \nonumber  \\
c_{2,k} &=& - \frac{1}{(k+1)(k+2)}, \qquad
d_{2,k} = \frac{k(k+6)}{(k+1)(k+5)}\lambda^2.
\end{eqnarray}
{\bf Skvortsov-Vasiliev formalism} Let us partially fix the gauge by
setting $\phi^{\alpha(3)} = 0$. Then from the equation for 
$\phi^{\alpha(3)}$ we obtain
$$
\Phi^{\alpha(3)} = - e_{\beta\dot\alpha} 
\phi^{\alpha(3)\beta\dot\alpha},
$$
while the equation for $\Phi^{\alpha(3)}$ takes the form
$$
e_{\beta\dot\alpha} D \phi^{\alpha(3)\beta\dot\alpha}
= E_{\beta(2)} Y^{\alpha(3)\beta(2)}
$$
that is completely consistent with the equations for the gauge
invariant zero-forms. This leaves us with the physical one-forms only
with the curvatures
\begin{eqnarray}
{\cal F}^{\alpha(2)\dot\alpha} &=& D \Phi^{\alpha(2)\dot\alpha}
+ \frac{M}{2} e^\alpha{}_{\dot\beta} \Phi^{\alpha\dot\alpha\dot\beta}
+ \frac{a_0}{3} e^{\alpha\dot\alpha} \Phi^\alpha, \nonumber \\
{\cal F}^\alpha &=& D \Phi^\alpha + a_0 e_{\beta\dot\alpha}
\Phi^{\alpha\beta\dot\alpha} + \frac{3M}{2} e^\alpha{}_{\dot\alpha}
\Phi^{\dot\alpha},
\end{eqnarray}
which are still invariant under $\rho^{\alpha(2)\dot\alpha}$ and 
$\rho^\alpha$ gauge transformations. Let us stress that 
${\cal F}^{\alpha(2)\dot\alpha}$ does not vanish on-shell now.
In this gauge the free Lagrangian is simply
\begin{equation}
{\cal L}_0 = - \frac{1}{2M} {\cal F}_{\alpha(2)\dot\alpha}
{\cal F}^{\alpha(2)\dot\alpha} + \frac{1}{3M} {\cal F}_\alpha
{\cal F}^\alpha + h.c. 
\end{equation}

\subsection{Massless spin 5/2}

Here we have just two physical helicities $(\pm \frac{5}{2})$ so we
introduce one-forms $\Phi^{\alpha(2)\dot\alpha} + h.c.$. Free
Lagrangian looks as follows:
\begin{equation}
{\cal L}_0 = \Phi_{\alpha\beta\dot\alpha} e^\beta{}_{\dot\beta}
D \Phi^{\alpha\dot\alpha\dot\beta} + \lambda 
[ 3 \Phi_{\alpha\beta\dot\alpha} E^\beta{}_\gamma
\Phi^{\alpha\gamma\dot\alpha} - \Phi_{\alpha(2)\dot\alpha}
E^{\dot\alpha}{}_{\dot\beta} \Phi^{\alpha(2)\dot\beta}] + h.c. 
\end{equation}
The Lagrangian is invariant under the following local transformations:
\begin{equation}
\delta \Phi^{\alpha(2)\dot\alpha} = D \rho^{\alpha(2)\dot\alpha}
+ e_\beta{}^{\dot\alpha} \rho^{\alpha(2)\beta} + \lambda
e^\alpha{}_{\dot\beta} \rho^{\alpha\dot\alpha\dot\beta}.
\end{equation}
Once again we see gauge transformations with parameter 
$\rho^{\alpha(2)\dot\alpha}$ and algebraic one with parameter 
$\rho^{\alpha(3)}$. So we introduce an extra field $\Phi^{\alpha(3)}$
with gauge transformations
\begin{equation}
\delta \Phi^{\alpha(3)} = D \rho^{\alpha(3)} + \lambda^2
e^\alpha{}_{\dot\alpha} \rho^{\alpha(2)\dot\alpha}. 
\end{equation}
Then we obtain a complete set of gauge invariant curvatures
\begin{eqnarray}
{\cal F}^{\alpha(3)} &=& D \Phi^{\alpha(3)} + \lambda^2
e^\alpha{}_{\dot\alpha} \Phi^{\alpha(2)\dot\alpha}, \nonumber \\
{\cal F}^{\alpha(2)\dot\alpha} &=& D \Phi^{\alpha(2)\dot\alpha}
+ e_\beta{}^{\dot\alpha} \Phi^{\alpha(2)\beta} + \lambda
e^\alpha{}_{\dot\beta} \Phi^{\alpha\dot\alpha\dot\beta}.
\end{eqnarray}
Note that on-shell we have
\begin{equation}
{\cal F}^{\alpha(2)\dot\alpha} \approx 0. 
\end{equation}
The Lagrangian in terms of curvatures has the form:
\begin{equation}
{\cal L}_0 = - \frac{1}{12\lambda^3} {\cal F}_{\alpha(3)} 
{\cal F}^{\alpha(3)} - \frac{1}{4\lambda} 
{\cal F}_{\alpha(2)\dot\alpha} {\cal F}^{\alpha(2)\dot\alpha} + h.c.
\end{equation}

\subsection{Partially massless spin 2}

Partially massless spin 2 has four physical helicities 
$(\pm 2, \pm 1)$ so we introduce one-forms 
$\Omega^{\alpha(2)} + h.c.$, $H^{\alpha\dot\alpha}$,
$A$ and zero-form $B^{\alpha(2)} + h.c.$. The free Lagrangian 
looks like:
\begin{eqnarray}
{\cal L}_0 &=& \Omega^{\alpha\beta} E_\beta{}^\gamma
\Omega_{\alpha\gamma} + \Omega^{\alpha\beta} e_\beta{}^{\dot\alpha}
D H_{\alpha\dot\alpha} + 4 E B_{\alpha(2)} B^{\alpha(2)} + 2 
E_{\alpha(2)} B^{\alpha(2)} D A \nonumber \\
 && - 2m E_{\alpha(2)} \Omega^{\alpha(2)} A + 4m B^{\alpha\beta}
E_\beta{}^{\dot\alpha} H_{\alpha\dot\alpha} + h.c. 
\end{eqnarray}
It is invariant under the following gauge transformations
\begin{eqnarray}
\delta \Omega^{\alpha(2)} &=& D \eta^{\alpha(2)}, \qquad 
\delta H^{\alpha\dot\alpha} = D \xi^{\alpha\dot\alpha} + 
e_\beta{}^{\dot\alpha} \eta^{\alpha\beta} + e^\alpha{}_{\dot\beta}
\eta^{\dot\alpha\dot\beta} + m e^{\alpha\dot\alpha} \xi, \nonumber \\
\delta B^{\alpha(2)} &=& - m \eta^{\alpha(2)}, \qquad
\delta A = D \xi + m e^{\alpha\dot\alpha} \xi_{\alpha\dot\alpha}, 
\end{eqnarray}
where
$$
m^2 = - 2\lambda^2.
$$
As it common in the frame-like formalism each field (both physical as
well as auxiliary) has its own gauge invariant object (curvature):
\begin{eqnarray}
{\cal R}^{\alpha(2)} &=& D \Omega^{\alpha(2)} + m E^\alpha{}_\beta
B^{\alpha\beta}, \nonumber \\
{\cal T}^{\alpha\dot\alpha} &=& D H^{\alpha\dot\alpha} + 
e_\beta{}^{\dot\alpha} \Omega^{\alpha\beta} + e^\alpha{}_{\dot\beta}
\Omega^{\dot\alpha\dot\beta} + m e^{\alpha\dot\alpha} A, \\
{\cal A} &=& D A + 2(E^{\alpha(2)} B_{\alpha(2)} + E^{\dot\alpha(2)}
B_{\dot\alpha(2)}) + m e^{\alpha\dot\alpha} H_{\alpha\dot\alpha},
\nonumber \\
{\cal B}^{\alpha(2)} &=& D B^{\alpha(2)} + m \Omega^{\alpha(2)}.
\nonumber
\end{eqnarray}
By analogy with the frame formalism in (super)gravity we use a kind of
''zero torsion condition''\footnote{In this case it means that we work
''partially on-shell'', i.e. on the auxiliary field equations.}:
\begin{equation}
{\cal T}^{\alpha\dot\alpha} \approx 0, \qquad {\cal A} \approx 0.
\end{equation}
In this case the remaining curvatures satisfy a number of differential
identities:
\begin{equation}
D {\cal R}^{\alpha(2)} \approx m E^\alpha{}_\beta 
{\cal B}^{\alpha\beta}, \qquad D {\cal B}^{\alpha\beta} \approx m 
{\cal R}^{\alpha\beta},
\end{equation}
as well as some algebraic ones
\begin{equation}
e_\beta{}^{\dot\alpha} {\cal R}^{\alpha\beta} + e^\alpha{}_{\dot\beta}
{\cal R}^{\dot\alpha\dot\beta} \approx 0, \qquad
E_{\alpha(2)} {\cal B}^{\alpha(2)} + E_{\dot\alpha(2)}
{\cal B}^{\dot\alpha(2)} \approx 0. 
\end{equation}
The free Lagrangian can be rewritten as
\begin{equation}
{\cal L}_0 = d_1 {\cal R}^{\alpha(2)} {\cal R}_{\alpha(2)} + d_2
{\cal B}^{\alpha\beta} E_\beta{}^\gamma {\cal B}_{\alpha\gamma}
+ \frac{1}{m} {\cal B}^{\alpha\beta} e_\beta{}^{\dot\alpha}
{\cal T}_{\alpha\dot\alpha} + h.c. 
\end{equation}
where 
$$
2d_1 - d_2 = \frac{1}{m^2}.
$$

Now let us turn to the unfolded equations. For the gauge sector (i.e.
sector of gauge one-forms and Stueckelberg zero-form) we have
\begin{eqnarray}
0 &=& D \Omega^{\alpha(2)} + m E^\alpha{}_\beta B^{\alpha\beta}
+ E_{\beta(2)} W^{\alpha(2)\beta(2)}, \nonumber \\
0 &=& D H^{\alpha\dot\alpha} + e_\beta{}^{\dot\alpha}
\Omega^{\alpha\beta} + e^\alpha{}_{\dot\beta}
\Omega^{\dot\alpha\dot\beta} + m e^{\alpha\dot\alpha} A, \\
0 &=& D A + 2 (E_{\alpha(2)} \Omega^{\alpha(2)} + E_{\dot\alpha(2)}
\Omega^{\dot\alpha(2)}) + m e_{\alpha\dot\alpha} H^{\alpha\dot\alpha},
\nonumber \\
0 &=& D B^{\alpha(2)} + m \Omega^{\alpha(2)} + e_{\beta\dot\alpha}
B^{\alpha(2)\beta\dot\alpha}. \nonumber
\end{eqnarray}
Here $W^{\alpha(4)}$ and $B^{\alpha(3)\dot\alpha}$ are first
representatives of two infinite chains of gauge invariant zero-forms.
Their unfolded equations are $(k \ge 0)$:
\begin{eqnarray}
0 &=& D W^{\alpha(4+k)\dot\alpha(k)} + e_{\beta\dot\beta}
W^{\alpha(4+k)\beta\dot\alpha(k)\dot\beta} + a_{1,k}
e^\alpha{}_{\dot\beta} B^{\alpha(3+k)\dot\alpha(k)\dot\beta}
+ b_{1,k} e^{\alpha\dot\alpha} W^{\alpha(3+k)\dot\alpha(k-1)},
\nonumber \\
0 &=& D B^{\alpha(3+k)\dot\alpha(k+1)} + e_{\beta\dot\beta}
B^{\alpha(3+k)\beta\dot\alpha(k+1)\dot\beta} + a_{2,k}
e_\beta{}^{\dot\alpha} W^{\alpha(3+k)\beta\dot\alpha(k)} 
+ b_{2,k} e^{\alpha\dot\alpha} B^{\alpha(2+k)\dot\alpha(k)}, 
\end{eqnarray}
where
\begin{eqnarray}
a_{1,k} &=& \frac{4m}{(k+4)(k+5)}, \qquad
a_{2,k} = \frac{m}{(k+1)(k+2)}, \nonumber \\
b_{1,k} &=& \frac{(k+2)(k+3)}{(k+1)(k+4)}\lambda^2, \qquad
b_{2,k} = \frac{k(k+5)}{(k+1)(k+4)}\lambda^2.
\end{eqnarray}

\subsection{Massless spin 2}

We need one-forms $\Omega^{\alpha(2)} + h.c.$ and
$H^{\alpha\dot\alpha}$. The free Lagrangian
\begin{equation}
{\cal L}_0 = \Omega^{\alpha\beta} E_\beta{}^\gamma
\Omega_{\alpha\gamma} + \Omega^{\alpha\beta} e_\beta{}^{\dot\alpha}
D H_{\alpha\dot\alpha} + 2\lambda^2 H^{\alpha\dot\alpha} 
E_\alpha{}^\beta H_{\beta\dot\alpha} + h.c.
\end{equation}
is invariant under the gauge transformations
\begin{eqnarray*}
\delta \Omega^{\alpha(2)} &=& D \eta^{\alpha(2)} + \lambda^2 
e^\alpha{}_{\dot\alpha} \xi^{\alpha\dot\alpha}, \nonumber \\
\delta H^{\alpha\dot\alpha} &=& D \xi^{\alpha\dot\alpha} +
e_\beta{}^{\dot\alpha} \eta^{\alpha\beta} + e^\alpha{}_{\dot\beta}
\eta^{\dot\alpha\dot\beta}.
\end{eqnarray*}
Gauge invariant curvatures look like:
\begin{eqnarray}
{\cal R}^{\alpha(2)} &=& D \Omega^{\alpha(2)} + \lambda^2
e^\alpha{}_{\dot\alpha} H^{\alpha\dot\alpha}, \nonumber \\
{\cal T}^{\alpha\dot\alpha} &=& D H^{\alpha\dot\alpha} + 
e_\beta{}^{\dot\alpha} \Omega^{\alpha\beta} + e^\alpha{}_{\dot\beta}
\Omega^{\dot\alpha\dot\beta}. 
\end{eqnarray}
Here we also use zero-torsion condition:
\begin{equation}
{\cal T}^{\alpha\dot\alpha} \approx 0 \quad \Rightarrow \quad
e_\beta{}^{\dot\alpha} {\cal R}^{\alpha\beta} +
e^\alpha{}_{\dot\beta} {\cal R}^{\dot\alpha\dot\beta} \approx 0.
\end{equation}
The free Lagrangian has the form:
\begin{equation}
{\cal L}_0 = \frac{i}{4\lambda^2} {\cal R}_{\alpha(2)} 
{\cal R}^{\alpha(2)} + h.c.  
\end{equation}
Unfolded equations for the one-forms look like:
\begin{eqnarray}
0 &=&  D \Omega^{\alpha(2)} + \lambda^2 e^\alpha{}_{\dot\alpha}
H^{\alpha\dot\alpha} + E_{\beta(2)} W^{\alpha(2)\beta(2)},
\nonumber \\
0 &=& D H^{\alpha\dot\alpha} + e_\beta{}^{\dot\alpha}
\Omega^{\alpha\beta} + e^\alpha{}_{\dot\beta} 
\Omega^{\dot\alpha\dot\beta}, 
\end{eqnarray}
while for the gauge invariant zero-forms we have ($k \ge 0$):
\begin{equation}
0 = D W^{\alpha(4+k)\dot\alpha(k)} + e_{\beta\dot\beta}
W^{\alpha(4+k)\beta\dot\alpha(k)\dot\beta} + \lambda^2
e^{\alpha\dot\alpha} W^{\alpha(3+k)\dot\alpha(k-1)}.
\end{equation}

\subsection{Massless spin 3/2}

We need only one-form $\Psi^\alpha + h.c.$. The free Lagrangian looks
like
\begin{equation}
{\cal L}_0 = - \Psi_\alpha e^\alpha{}_{\dot\alpha} D \Psi^{\dot\alpha}
- \lambda \Psi_\alpha E^\alpha{}_\beta \Psi^\beta + h.c.
\end{equation}
Gauge transformations leaving this Lagrangian invariant are
\begin{equation}
\delta \Psi^\alpha = D \zeta^\alpha + \lambda
e^\alpha{}_{\dot\alpha} \zeta^{\dot\alpha},
\end{equation}
while the gauge invariant curvature looks like
\begin{equation}
\tilde{\cal F}^\alpha = D \Psi^\alpha + \lambda 
e^\alpha{}_{\dot\alpha} \Psi^{\dot\alpha}.
\end{equation}
It satisfies an identity
\begin{equation}
D \tilde{\cal F}^\alpha = - \lambda e^\alpha{}_{\dot\alpha}
\tilde{\cal F}^{\dot\alpha}. 
\end{equation}
The free Lagrangian can be written as
\begin{equation}
{\cal L}_0 = \frac{1}{4\lambda} \tilde{\cal F}_\alpha
\tilde{\cal F}^\alpha + h.c.
\end{equation}
Unfolded equations for gauge fields have the form
\begin{eqnarray}
0 &=& D \Psi^\alpha + \lambda e^\alpha{}_{\dot\alpha}
\Psi^{\dot\alpha} + E_{\beta(2)} Y^{\alpha\beta(2)}, \nonumber \\
0 &=& D \Psi^{\dot\alpha} + \lambda e_\alpha{}^{\dot\alpha}
\Psi^\alpha + E_{\dot\beta(2)} Y^{\dot\alpha\dot\beta(2)},
\end{eqnarray}
while equations for the gauge invariant zero-forms look like 
$(k \ge 0)$:
\begin{equation}
0 = D Y^{\alpha(3+k)\dot\alpha(k)} + e_{\beta\dot\beta}
Y^{\alpha(3+k)\beta\dot\alpha(k)\dot\beta} + \lambda^2
e^{\alpha\dot\alpha} Y^{\alpha(2+k)\dot\alpha(k-1)}. 
\end{equation}

\section{Warming up: massless supermultiplet $(5/2,2)$}

We begin with the constructive approach. We use a zero torsion
condition ${\cal T}^{\alpha\dot\alpha} \approx 0$ and consider
supertransformations for the physical fields only. Our ansatz (the
choice of coefficients will become clear later):
\begin{equation}
\delta \Phi^{\alpha(2)\dot\alpha} = a_2 H^{\alpha\dot\alpha}
\zeta^\alpha, \qquad \delta \Phi^{\alpha\dot\alpha(2)} = a_1
\Omega^{\dot\alpha(2)} \zeta^\alpha, \qquad
\delta H^{\alpha\dot\alpha} = b_2 \Phi^{\alpha\beta\dot\alpha}
\zeta_\beta + h.c.
\end{equation}
Variation of the massless spin 2 Lagrangian produces
\begin{equation}
\Delta _B = 2b_2 e_\beta{}^{\dot\alpha} \Phi_{\alpha\gamma\dot\alpha}
{\cal R}^{\alpha\beta} \zeta^\gamma + h.c., 
\end{equation}
while variation of the massless spin 5/2 Lagrangian gives
\begin{equation}
\Delta_F = - a_1 e^\beta{}_{\dot\alpha} 
{\cal F}_{\alpha\beta\dot\alpha} \Omega^{\dot\alpha(2)} \zeta^\alpha
+ a_2 e_\alpha{}^{\dot\beta} {\cal F}_{\alpha\dot\alpha\dot\beta}
H^{\alpha\dot\alpha} \zeta^\alpha.
\end{equation}
Now using explicit expression for the spin 5/2 curvature and
integrating by parts one can show that at $a_2 = \lambda a_1$ this
variation can  be rewritten in the on-shell equivalent form:
\begin{eqnarray}
\Delta_F &=& 2a_1 e_\alpha{}^{\dot\alpha}
\Phi_{\alpha\gamma\dot\alpha} {\cal R}^{\alpha\beta} \zeta^\gamma
\nonumber \\
 && + 2a_1 [ e^\beta{}_{\dot\alpha} \Phi_{\alpha\beta\dot\beta}
\Omega^{\dot\alpha\dot\beta} - \lambda e_{(\alpha}{}^{\dot\alpha}
\Phi_{\beta)\dot\alpha\dot\beta} H^{\beta\dot\beta}]
[ D \zeta^\alpha + \lambda e^\alpha{}_{\dot\gamma} \zeta^{\dot\gamma}]
+ h.c. 
\end{eqnarray}
At $a_1 = - b_2$ the first term compensates bosonic variation while
the second line gives us a supercurrent so that to achieve invariance
under the supertransformations we have to introduce a cubic vertex
\begin{equation}
{\cal L}_1 = - 2a_1 [ e^\beta{}_{\dot\alpha}
\Phi_{\alpha\beta\dot\beta} \Omega^{\dot\alpha\dot\beta} - \lambda
e_{(\alpha}{}^{\dot\alpha} \Phi_{\beta)\dot\alpha\dot\beta}
H^{\beta\dot\beta}] \Psi^\alpha + h.c. 
\end{equation}
But it is not the end of the story. Both spin 5/2 and spin 2 are gauge
fields themselves so the correct vertex must also be invariant under
their gauge transformations, Therefore, we must find appropriate
non-minimal terms\footnote{Here non-minimal means that gravitino must
enter only through its gauge invariant field strength.}. To find a
complete gauge invariant vertex let us turn to the Fradkin-Vasiliev
formalism.

The first step is to construct consistent deformations for all gauge
invariant curvatures. Here the consistency means that the deformed
curvatures transform covariantly.\\
{\bf Spin 5/2} The most general ansatz looks like:
\begin{eqnarray}
\Delta {\cal F}^{\alpha(3)} &=& a_0 \Omega^{\alpha(2)} \Psi^\alpha,
\nonumber \\
\Delta {\cal F}^{\alpha(2)\dot\alpha} &=& a_1 \Omega^{\alpha(2)}
\Psi^{\dot\alpha} + a_2 H^{\alpha\dot\alpha} \Psi^\alpha, 
\end{eqnarray}
and implies the following supertransformations:
\begin{equation}
\delta \Phi^{\alpha(3)} = a_0 \Omega^{\alpha(2)} \zeta^\alpha, \qquad
\delta \Phi^{\alpha(2)\dot\alpha} = a_2 H^{\alpha\dot\alpha}
\zeta^\alpha, \qquad \delta \Phi^{\alpha\dot\alpha(2)} = a_1
\Omega^{\dot\alpha(2)} \zeta^\alpha. 
\end{equation}
Then for the variations of the deformed curvatures we obtain
\begin{eqnarray*}
\delta \hat{\cal F}^{\alpha(3)} &=& a_0 D \Omega^{\alpha(2)}
\zeta^\alpha + \lambda^2 a_2 e^\alpha{}_{\dot\alpha}
H^{\alpha\dot\alpha}, \\
\delta \hat{\cal F}^{\alpha(2)\dot\alpha} &=& a_2 D
H^{\alpha\dot\alpha} \zeta^\alpha + (a_0 - \lambda a_1) 
e_\beta{}^{\dot\alpha} \Omega^{\alpha(2)} \zeta^\beta + a_0 
e_\beta{}^{\dot\alpha} \Omega^{\alpha\beta} \zeta^\alpha
+ \lambda a_1 e^\alpha{}_{\dot\beta} \Omega^{\dot\alpha\dot\beta}
\zeta^\alpha.
\end{eqnarray*}
Thus if we put
$$
a_0 = \lambda a_1 = a_2,
$$
we obtain
\begin{equation}
\delta \hat{\cal F}^{\alpha(3)} =  a_0 {\cal R}^{\alpha\beta}
\zeta^\alpha, \qquad \delta \hat{\cal F}^{\alpha(2)\dot\alpha}
= a_2 {\cal T}^{\alpha\dot\alpha} \zeta^\alpha.
\end{equation}
It is easy to check that the deformed curvatures transform
covariantly also under the spin 2 gauge transformations
\begin{eqnarray}
\delta \hat{\cal F}^{\alpha(3)} &=& - a_0 \eta^{\alpha(2)}
\tilde{\cal F}^\alpha, \nonumber \\
\delta \hat{\cal F}^{\alpha(2)\dot\alpha} &=& - a_1 \eta^{\alpha(2)}
\tilde{\cal F}^{\dot\alpha} - a_2 \xi^{\alpha\dot\alpha}
\tilde{\cal F}^\alpha.
\end{eqnarray}
{\bf Spin 2} An ansatz looks like
\begin{eqnarray}
\Delta {\cal R}^{\alpha(2)} &=& b_0 \Phi^{\alpha(2)\beta} \Psi_\beta +
b_1 \Phi^{\alpha(2)\dot\alpha} \Psi_{\dot\alpha}, \nonumber \\
\Delta {\cal T}^{\alpha\dot\alpha} &=& b_2
\Phi^{\alpha\beta\dot\alpha} \Psi_\beta + h.c.,
\end{eqnarray}
and leads to the following supertransformations
\begin{equation}
\delta \Omega^{\alpha(2)} = b_0 \Phi^{\alpha(2)\beta} \zeta_\beta,
\qquad \delta \Omega^{\dot\alpha(2)} = b_1 \Phi^{\alpha\dot\alpha(2)}
\Psi_\alpha, \qquad \delta H^{\alpha\dot\alpha} = b_2
\Phi^{\alpha\beta\dot\alpha} \zeta_\beta.
\end{equation}
If we put
$$
b_1 = \lambda b_0, \qquad b_2 = b_0
$$
we obtain
\begin{eqnarray}
\delta \hat{\cal R}^{\alpha(2)} &=& b_0 {\cal F}^{\alpha(2)\beta}
\zeta_\beta + b_1 {\cal F}^{\alpha(2)\dot\alpha} \zeta_{\dot\alpha},
\nonumber \\
\delta \hat{\cal T}^{\alpha\dot\alpha} &=& b_2 
{\cal F}^{\alpha\beta\dot\alpha} \zeta_\beta. 
\end{eqnarray}
At the same time for the hypertransformations we have
\begin{eqnarray}
\delta \hat{\cal R}^{\alpha(2)} &=& - b_0 \rho^{\alpha(2)\beta}
\tilde{\cal F}_\beta - b_1 \rho^{\alpha(2)\dot\alpha}
\tilde{\cal F}_{\dot\alpha}, \nonumber \\
\delta \hat{\cal T}^{\alpha\dot\alpha} &=& - b_2
\rho^{\alpha\beta\dot\alpha} \tilde{\cal F}_\beta + h.c.
\end{eqnarray}
{\bf Spin 3/2} An ansatz has the form
\begin{equation}
\Delta \tilde{\cal F}^\alpha = c_0 \Phi^{\alpha\beta(2)}
\Omega_{\beta(2)} + c_1 \Phi^{\alpha\dot\alpha(2)}
\Omega_{\dot\alpha(2)} + c_2 \Phi^{\alpha\beta\dot\alpha}
H_{\beta\dot\alpha}. 
\end{equation}
Here consistency requires
$$
c_1 = \lambda c_0, \qquad c_2 = 2\lambda^2 c_0 ,
$$
and we have
\begin{eqnarray}
\delta \hat{\tilde{\cal F}}^\alpha &=& c_0 {\cal F}^{\alpha\beta(2)}
\eta_{\beta(2)} + c_1 {\cal F}^{\alpha\dot\alpha(2)}
\eta_{\dot\alpha(2)} + c_2 {\cal F}^{\alpha\beta\dot\alpha}
\xi_{\beta\dot\alpha} \nonumber \\
 && - c_0 \rho^{\alpha\beta(2)} {\cal R}_{\beta(2)} - c_1
\rho^{\alpha\dot\alpha(2)} {\cal R}_{\dot\alpha(2)} - c_2
\rho^{\alpha\beta\dot\alpha} {\cal T}_{\beta\dot\alpha}.
\end{eqnarray}

Now we consider a deformed Lagrangian (i.e. the sum of the free
Lagrangians where all curvatures are replaced with the deformed ones)
and require it to be invariant under all gauge transformations.
For the variations of the deformed Lagrangian that do not vanish 
on-shell we obtain
\begin{eqnarray}
\delta \hat{\cal L} &\approx& [ - \frac{a_0}{2\lambda^3} - 
\frac{b_0}{2\lambda^2}]   {\cal F}_{\alpha(2)\beta} 
{\cal R}^{\alpha(2)} \zeta^\beta  + [ \frac{a_0}{2\lambda^3} +
\frac{c_0}{2\lambda}] {\cal F}_{\alpha(2)\beta} \eta^{\alpha(2)}
\tilde{\cal F}^\alpha  + [ \frac{b_0}{2\lambda^2} -
\frac{c_0}{2\lambda} ] \rho_{\alpha(2)\beta} {\cal R}^{\alpha(2)}
\tilde{\cal F}^\beta \nonumber \\
 && [ - \frac{a_1}{2\lambda} - \frac{b_1}{2\lambda^2}]
 {\cal F}_{\alpha(2)\dot\alpha} {\cal R}^{\alpha(2)}
\zeta^{\dot\alpha} + [ \frac{a_2}{\lambda} + \frac{c_2}{2\lambda} ]
{\cal F}_{\alpha\beta\dot\alpha} \xi^{\alpha\dot\alpha} 
\tilde{\cal F}^\beta. 
\end{eqnarray}
This gives
$$
\lambda b_0 = \lambda^2 c_0 = - a_0.
$$
At last we extract the cubic part of the deformed Lagrangian
\begin{eqnarray}
{\cal L}_1 &=& \frac{c_0}{2\lambda} {\cal F}_{\alpha(2)\beta}
 \Omega^{\alpha(2)} \Psi^\beta + \frac{1}{2} 
{\cal F}_{\alpha(2)\dot\alpha}  [ c_0 \Omega^{\alpha(2)}
\Psi^{\dot\alpha} + \lambda c_0 H^{\alpha\dot\alpha} \Psi^\alpha ]
\nonumber \\
 && - \frac{1}{2\lambda} [ c_0 \Phi_{\alpha(2)\beta}
{\cal R}^{\alpha(2)} \Psi^\beta + \lambda c_0 
\Phi_{\alpha(2)\dot\alpha} {\cal R}^{\alpha(2)} \Psi^{\dot\alpha} ]
\nonumber \\
 && + \frac{1}{2\lambda} [ c_0 \Phi_{\alpha\beta(2)} 
\Omega^{\beta(2)} + \lambda c_0 \Phi_{\alpha\dot\alpha(2)} 
\Omega^{\dot\alpha(2)} + 2\lambda^2 c_0 \Phi_{\alpha\beta\dot\alpha}
H^{\beta\dot\alpha}] \tilde{\cal F}^\alpha.
\end{eqnarray}
This vertex contains terms with up to three derivatives but the terms
with three derivatives combine into total derivative and can be
dropped (as it is usual to be the case for the massless fields, see
\cite{Vas11,KhZ20a}). The remaining part (up to total derivatives and
vanishing on-shell terms) can be rewritten in the following simple
form
\begin{eqnarray}
{\cal L}_1 &=& - c_0 \Phi_{\alpha(2)\dot\alpha} \Omega^{\alpha(2)}
[ D \Psi^{\dot\alpha} + \lambda e_\beta{}^{\dot\alpha} \Psi^\beta]
\nonumber \\
 && + 2\lambda c_0 [ e^\beta{}_{\dot\alpha}
\Phi_{\alpha\beta\dot\beta} \Omega^{\dot\alpha\dot\beta} - \lambda 
e_{(\alpha}{}^{\dot\alpha} \Phi_{\beta)\dot\alpha\dot\beta}
H^{\beta\dot\beta} ] \Psi^\alpha. 
\end{eqnarray}
In the first line we see the non-minimal interaction that is
necessary for the vertex to be invariant under all gauge
transformations, while in the second line we see the supercurrent we
already familiar with.

\section{Massless spin 2}

In this Section we consider massless spin 2 as a superpartner for the
partially massless spin 5/2. We elaborate on both possible description
for partially massless spin 5/2.

\subsection{Skvortsov-Vasiliev formalism}

Here we use our analogue of the Skvortsov-Vasiliev formalism for the
partially massless spin 5/2. This formalism does not contain any
Stueckelberg zero-forms. As a result, there are no any ambiguities
related with the field redefinitions and the Fradkin-Vasiliev
formalism works exactly as in the massless case. We begin with the
construction of consistent deformations for all gauge invariant
curvatures. \\
{\bf Partially massless spin 5/2} The most general ansatz looks like:
\begin{eqnarray}
\Delta {\cal F}^{\alpha(2)\dot\alpha} &=& a_1 \Omega^{\alpha(2)}
\Psi^{\dot\alpha} + a_2 H^{\alpha\dot\alpha} \Psi^\alpha, \nonumber \\
\Delta {\cal F}^\alpha &=& a_3 \Omega^{\alpha\beta} \Psi_\beta + a_4
H^{\alpha\dot\alpha} \Psi_{\dot\alpha}, 
\end{eqnarray}
and its consistency requires:
\begin{equation}
M = - \lambda, \qquad a_2 = - \frac{\lambda}{2}a_1, \qquad
a_3 = \frac{2a_0}{5\lambda}a_1, \qquad a_4 = - \frac{3a_0}{5}a_1.
\end{equation}
Variations of the deformed curvatures
\begin{eqnarray}
\delta \hat{\cal F}^{\alpha(2)\dot\alpha} &=& a_1 {\cal R}^{\alpha(2)}
\zeta^{\dot\alpha} + a_2 {\cal T}^{\alpha\dot\alpha} \zeta^\alpha 
- a_1 \eta^{\alpha(2)} \tilde{\cal F}^{\dot\alpha} - a_2
\xi^{\alpha\dot\alpha} \tilde{\cal F}^\alpha, \nonumber \\
\delta \hat{\cal F}^\alpha &=& a_3 {\cal R}^{\alpha\beta} \zeta_\beta
+ a_4 {\cal T}^{\alpha\dot\alpha} \zeta_{\dot\alpha} - a_3
\eta^{\alpha\beta} \tilde{\cal F}_\beta - a_4 \xi^{\alpha\dot\alpha}
\tilde{\cal F}_{\dot\alpha}, 
\end{eqnarray}
show that the deformed curvatures do transform covariantly. \\
{\bf Massless spin 2} Here an ansatz has the form:
\begin{eqnarray}
\Delta {\cal R}^{\alpha(2)} &=& b_2 \Phi^{\alpha(2)\dot\alpha}
\Psi_{\dot\alpha} + b_3 \Phi^\alpha \Psi^\alpha, \nonumber\\
\Delta {\cal T}^{\alpha\dot\alpha} &=& b_5
(\Phi^{\alpha\beta\dot\alpha} \Psi_\beta +
\Phi^{\alpha\dot\alpha\dot\beta} \Psi_{\dot\beta}) + b_6
(\Phi^\alpha \Psi^{\dot\alpha} + \Phi^{\dot\alpha} \Psi^\alpha),
\end{eqnarray}
and for consistency we must put
\begin{equation}
M = - \lambda, \quad
b_3 = - \frac{\lambda}{2a_0}b_2, \quad 
b_5 = - \frac{1}{2\lambda}b_2, \quad
b_6 = \frac{3}{4a_0}b_2.
\end{equation}
Variations of the deformed curvatures are also covariant:
\begin{eqnarray}
\delta \hat{\cal R}^{\alpha(2)} &=& b_2 {\cal F}^{\alpha(2)\dot\alpha}
\zeta_{\dot\alpha} + b_3 {\cal F}^\alpha \zeta^\alpha - b_2
\rho^{\alpha(2)\dot\alpha} \tilde{\cal F}_{\dot\alpha} - b_3
\rho^\alpha \tilde{\cal F}^\alpha, \nonumber \\
\delta \hat{\cal T}^{\alpha\dot\alpha} &=& b_5 
{\cal F}^{\alpha\beta\dot\alpha} \zeta_\beta + b_6 {\cal F}^\alpha
\zeta^{\dot\alpha} - b_5 \rho^{\alpha\beta\dot\alpha}
\tilde{\cal F}_\beta - b_6 \rho^\alpha \tilde{\cal F}^{\dot\alpha}
+ h.c. 
\end{eqnarray}
{\bf Massless spin 3/2} At lat, for gravitino we take:
\begin{eqnarray}
\Delta \tilde{\cal F}^\alpha &=& c_2 \Phi^{\alpha\dot\alpha(2)}
\Omega_{\dot\alpha(2)} + c_3 \Phi^{\alpha\beta\dot\alpha} 
H_{\beta\dot\alpha} + c_4 \Phi_\beta \Omega^{\alpha\beta} + c_5
\Phi_{\dot\alpha} H^{\alpha\dot\alpha}. 
\end{eqnarray}
Here the solution is also unique:
\begin{equation}
c_3 = - \lambda c_2, \qquad
a_0c_4 = - \lambda c_2, \qquad
c_5 = - \frac{2a_0}{5}c_2.
\end{equation}

Variations of the deformed curvature look like:
\begin{eqnarray}
\delta \hat{\tilde{\cal F}}^\alpha &=& c_2 
{\cal F}^{\alpha\dot\alpha(2)} \eta_{\dot\alpha(2)} + c_3
{\cal F}^{\alpha\beta\dot\alpha} \xi_{\beta\dot\alpha} + c_4
\eta^{\alpha\beta} {\cal F}_\beta + c_5 \xi^{\alpha\dot\alpha}
{\cal F}_{\dot\alpha} \nonumber \\
 && - c_2 \rho^{\alpha\dot\alpha(2)} {\cal R}_{\dot\alpha(2)} - c_3
\rho^{\alpha\beta\dot\alpha} {\cal T}_{\beta\dot\alpha} - c_4
{\cal R}^{\alpha\beta} \rho_\beta - c_5 {\cal T}^{\alpha\dot\alpha}
\rho_{\dot\alpha}. 
\end{eqnarray}
As a next step, we consider a deformed Lagrangian $\hat{\cal L}$
(i.e. the sum of the three Lagrangians where all curvatures are
replaced with the deformed ones) and calculate all its
non-vanishing on-shell variations:
\begin{eqnarray}
\delta \hat{\cal L} &=& 
[ \frac{a_1}{\lambda} - \frac{b_2}{2\lambda^2}]
{\cal F}_{\alpha(2)\dot\alpha} {\cal R}^{\alpha(2)} \zeta^{\dot\alpha}
 + [ a_1 - \frac{c_2}{2}]
 {\cal F}_{\alpha\beta\dot\alpha} \xi^{\alpha\dot\alpha}
\tilde{\cal F}^\beta 
 + [  \frac{b_2}{2\lambda a_0} - \frac{c_2}{2a_0}]
 \rho_\alpha {\cal R}^{\alpha\beta} \tilde{\cal F}_\beta \nonumber \\
 && [  - \frac{2a_3}{3\lambda} + \frac{b_3}{\lambda^2}  ]
{\cal F}_\alpha {\cal R}^{\alpha\beta} \zeta_\beta 
 + [ \frac{2a_3}{3\lambda} - \frac{c_4}{2\lambda} ]
{\cal F}_\alpha \eta^{\alpha\beta} \tilde{\cal F}_\beta.
\end{eqnarray}
Thus for the Lagrangian to be gauge invariant we must put:
\begin{equation}
2\lambda a_1 = b_2 = \lambda c_2.
\end{equation}
As the last step we extract a cubic vertex
\begin{eqnarray}
{\cal L}_1 &=& \frac{1}{\lambda} {\cal F}_{\alpha(2)\dot\alpha}
[ a_1 \Omega^{\alpha(2)} \Psi^{\dot\alpha} - \frac{\lambda}{2}a_1
H^{\alpha\dot\alpha} \Psi^\alpha ] \nonumber \\
 && + {\cal F}_\alpha 
[ \frac{a_1}{a_0} \Omega^{\alpha\beta} \Psi_\beta +
\frac{2a_0}{5\lambda}a_1 H^{\alpha\dot\alpha} \Psi_{\dot\alpha} ]
\nonumber \\
 && + \frac{1}{\lambda} [ - a_1 \Phi_{\alpha(2)\dot\alpha} 
{\cal R}^{\alpha(2)} \Psi^{\dot\alpha} - \frac{\lambda}{a_0}a_1
\Phi_\alpha {\cal R}^{\alpha\beta} \Psi_\beta] \nonumber \\
 && + \frac{1}{\lambda} [  a_1
\Phi_{\alpha\dot\alpha(2)} \Omega^{\dot\alpha(2)} - \lambda a_1
\Phi_{\alpha\beta\dot\alpha} H^{\beta\dot\alpha} +
\frac{\lambda}{a_0}a_1 \Phi^\beta \Omega_{\alpha\beta} +
\frac{2a_0}{5}a_1 \Phi^{\dot\alpha} H_{\alpha\dot\alpha} ] 
\tilde{\cal F}^\alpha. 
\end{eqnarray}
Note that there are no terms with three derivatives. Up to total
derivatives and terms that vanish on-shell this vertex can be
equivalently written as
\begin{eqnarray}
{\cal L}_1 &=& - \frac{2a_1}{\lambda} \Phi_{\alpha(2)\dot\alpha}
\Omega^{\alpha(2)} \tilde{\cal F}^{\dot\alpha} -
\frac{3\lambda}{2a_0}a_1 \Phi_{\dot\alpha} H^{\alpha\dot\alpha}
\tilde{\cal F}_\alpha \nonumber \\
 && + 2a_1 \Phi_{\alpha\beta\dot\alpha} e^\alpha{}_{\dot\beta}
\Omega^{\dot\alpha\dot\beta} \Psi^\beta - 2\lambda a_1 
e_{(\alpha}{}^{\dot\alpha} \Phi_{\beta)\dot\alpha\dot\beta}
H^{\alpha\dot\beta} \Psi^\beta - \frac{\lambda}{2}a_1 
e^{\alpha\dot\alpha} \Phi_{\alpha\dot\alpha\dot\beta}
H^{\beta\dot\beta} \Psi_\beta  \nonumber \\
 && - \frac{3\lambda}{2a_0}a_1 \Phi_{\dot\alpha} 
[e_\alpha{}^{\dot\alpha} \Omega^{\alpha\beta} +
\Omega^{\dot\alpha\dot\beta} e^\beta{}_{\dot\beta}] \Psi_\beta 
+ \frac{a_0}{10}a_1 \Phi_\alpha e^{(\alpha}{}_{\dot\alpha}
H^{\beta)\dot\alpha} \Psi_\beta + \frac{a_0}{2}a_1 \Phi_\alpha
e_{\beta\dot\alpha} H^{\beta\dot\alpha} \Psi^\alpha. 
\end{eqnarray}
Note that the first term in the first line and the first two terms in
the second line are the same as in the massless spin 5/2 case.
Naturally, this vertex is not invariant under the $\rho^{\alpha(3)}$
transformations:
\begin{equation}
\delta {\cal L}_1 = \frac{2a_1}{\lambda} \rho_{\alpha(2)\beta}
e^\beta{}_{\dot\alpha} \Omega^{\alpha(2)} \tilde{\cal F}^{\dot\alpha}
+ \lambda a_1 \rho_{\alpha(2)\beta} E^{\alpha(2)} H^{\beta\dot\alpha}
\Psi_{\dot\alpha} + h.c. \label{noninv}
\end{equation}
In the next subsection we will see how this non-invariance is
compensated in the general formalism.

\subsection{General formalism}

Here we also begin with the construction of the consistent curvatures
deformations. \\
{\bf Partially massless spin 5/2} In this case the result is
practically the same as in the previous subsection:
\begin{eqnarray}
\Delta {\cal F}^{\alpha(3)} &=& 0, \qquad
\Delta {\cal C}^{\alpha(3)} = 0, \nonumber \\
\Delta {\cal F}^{\alpha(2)\dot\alpha} &=& a_1 \Omega^{\alpha(2)}
\Psi^{\dot\alpha} + a_2 H^{\alpha\dot\alpha} \Psi^\alpha, \\
\Delta {\cal F}^\alpha &=& a_3 \Omega^{\alpha\beta} \Psi_\beta + a_4
H^{\alpha\dot\alpha} \Psi_{\dot\alpha}, \nonumber
\end{eqnarray}
with
\begin{equation}
M = - \lambda, \qquad a_2 = - \frac{\lambda}{2}a_1, \qquad
a_3 = \frac{2a_0}{5\lambda}a_1, \qquad a_4 = - \frac{3a_0}{5}a_1.
\end{equation}
{\bf Massless spin 2} The most general ansatz looks like:
\begin{eqnarray}
\Delta {\cal R}^{\alpha(2)} &=& b_1 \Phi^{\alpha(2)\beta} \Psi_\beta
+ b_2 \Phi^{\alpha(2)\dot\alpha} \Psi_{\dot\alpha} + b_3 \Phi^\alpha
\Psi^\alpha + b_4 e_{\beta\dot\alpha} \phi^{\alpha(2)\beta}
\Psi^{\dot\alpha}, \nonumber\\
\Delta {\cal T}^{\alpha\dot\alpha} &=& b_5
\Phi^{\alpha\beta\dot\alpha} \Psi_\beta + b_6 \Phi^\alpha
\Psi^{\dot\alpha} + b_7 e_\beta{}^{\dot\alpha}
\phi^{\alpha\beta\gamma} \Psi_\gamma + h.c., 
\end{eqnarray}
and the general solution has two free parameters (we take $b_{1,2}$):
\begin{equation}
b_3 = - \frac{\lambda}{2a_0}b_2, \quad 
b_4 = b_2 - \lambda b_1, \quad
b_5 = - \frac{1}{2\lambda}b_2, \quad
b_6 = \frac{3}{4a_0}b_2, \quad
b_7 = b_1 + \frac{1}{2\lambda}b_2.
\end{equation}
But we must take into account that there exists a field redefinition
\begin{equation}
\Omega^{\alpha(2)} \Rightarrow \Omega^{\alpha(2)} + \kappa_1
\phi^{\alpha(2)\beta} \Psi_\beta. 
\end{equation}
This produces
\begin{eqnarray}
\Delta {\cal R}^{\alpha(2)} &=& [ D \phi^{\alpha(2)\beta} -
\Phi^{\alpha(2)\beta}] \Psi_\beta + \phi^{\alpha(2)\beta}
[ D \Psi_\beta - \lambda e_\beta{}^{\dot\alpha} \Psi_{\dot\alpha}]
\nonumber \\
 && + \Phi^{\alpha(2)\beta} \Psi_\beta + \lambda \phi^{\alpha(2)\beta}
e_\beta{}^{\dot\alpha} \Psi_{\dot\alpha}, \\
\Delta {\cal T}^{\alpha\dot\alpha} &=& e_\beta{}^{\dot\alpha}
\phi^{\alpha\beta\gamma} \Psi_\gamma + e^\alpha{}_{\dot\beta}
\phi^{\dot\alpha\dot\beta\dot\gamma} \Psi_{\dot\gamma}. \nonumber
\end{eqnarray}
Therefore, it makes shifts
\begin{equation}
\Delta b_1 = \kappa_1, \qquad
\Delta b_4 = - \lambda \kappa_1, \qquad
\Delta b_7 = \kappa_1
\end{equation}
and generates abelian vertices
\begin{equation}
\Delta {\cal L}_a \sim {\cal R}_{\alpha(2)} 
[ {\cal C}^{\alpha(2)\beta} \Psi_\beta + \phi^{\alpha(2)\beta}
\tilde{\cal F}_\beta ].
\end{equation}
Recall that field redefinitions generate combinations of abelian and
non-abelian vertices that vanish on-shell (up to total derivatives)
and so do not change on-shell part of the cubic vertex. Thus we have
only one non-trivial parameter and we use $b_2$ in what follows. 
Note also that the solution found in the Appendix A.1 corresponds to 
$b_7 = 0$.\\
{\bf Massless spin 3/2} ere the most general ansatz can written as
\begin{eqnarray}
\Delta \tilde{\cal F}^\alpha &=& c_1 \Phi^{\alpha\beta(2)}
\Omega_{\beta(2)} + c_2 \Phi^{\alpha\dot\alpha(2)}
\Omega_{\dot\alpha(2)} + c_3 \Phi^{\alpha\beta\dot\alpha} 
H_{\beta\dot\alpha} + c_4 \Phi_\beta \Omega^{\alpha\beta} + c_5
\Phi_{\dot\alpha} H^{\alpha\dot\alpha} \nonumber \\
 && + c_6 \phi^{\alpha\beta\gamma} e_\beta{}^{\dot\alpha}
H_{\gamma\dot\alpha} + c_7 e^\alpha{}_{\dot\alpha}
\phi^{\dot\alpha\dot\beta(2)} \Omega_{\dot\beta(2)}.
\end{eqnarray}
Here the general solution also contains two free parameters (we take
$c_{1,2}$)
\begin{equation}
c_3 = - \lambda c_2, \quad
a_0c_4 = - \lambda c_2, \quad
c_5 = - \frac{2a_0}{5}c_2, \quad
c_6 = 2\lambda^2c_1 + \lambda c_2, \quad
c_7 = \lambda c_1 - c_2.
\end{equation}
But here there exist a field redefinition
\begin{equation}
\Psi^\alpha \Rightarrow \Psi^\alpha + \kappa_2 \phi^{\alpha\beta(2)}
\Omega_{\beta(2)}. 
\end{equation}
It produces shifts
\begin{equation}
\Delta c_1 = \kappa_2, \qquad
\Delta c_6 = 2\lambda^2\kappa_2, \qquad
\Delta c_7 = \lambda \kappa_2
\end{equation}
and generates abelian vertices
\begin{equation}
\Delta {\cal L}_a \sim \tilde{\cal F}_\alpha 
[ {\cal C}^{\alpha\beta(2)} \Omega_{\beta(2)} + \phi^{\alpha\beta(2)}
{\cal R}_{\beta(2)} ].
\end{equation}
Therefore we have just one nontrivial parameter and we  use $c_2$ in
what follows.

Now let us consider a combination of the deformed Lagrangian and
abelian vertices:
$$
{\cal L} = \hat{\cal L}_0 + {\cal L}_a,
$$
where
\begin{equation}
{\cal L}_a = g_1 {\cal C}^{\alpha(2)\beta} {\cal R}_{\alpha(2)}
\Psi_\beta + g_2 \phi^{\alpha(2)\beta} {\cal R}_{\alpha(2)}
\tilde{\cal F}_\beta + g_3 {\cal C}^{\alpha(2)\beta}
\Omega_{\alpha(2)} \tilde{\cal F}_\beta + h.c.
\end{equation}
To simplify presentation we consider different transformations
separately.\\
{\bf Supersymmetry} Here non-trivial deformations look like:
\begin{eqnarray}
\delta \hat{\cal F}^{\alpha(2)\dot\alpha} &=& a_1 {\cal R}^{\alpha(2)}
\zeta^{\dot\alpha}, \qquad \delta \hat{\cal F}^\alpha = a_3
{\cal R}^{\alpha\beta} \zeta_\beta, \nonumber \\
\delta \hat{\cal R}^{\alpha(2)} &=& b_1 {\cal F}^{\alpha(2)\beta}
\zeta_\beta + b_2 {\cal F}^{\alpha(2)\dot\alpha} \zeta_{\dot\alpha}
 + b_3 {\cal F}^\alpha \zeta^\alpha - b_4 e_{\beta\dot\alpha}
{\cal C}^{\alpha(2)\beta} \zeta^{\dot\alpha}.
\end{eqnarray}
Up to the terms that vanish on-shell we obtain
\begin{eqnarray}
\delta {\cal L} &=& [\frac{a_1}{\lambda} - \frac{b_2}{2\lambda^2}] 
{\cal F}_{\alpha(2)\dot\alpha} {\cal R}^{\alpha(2)} \zeta^{\dot\alpha}
+ [\frac{b_3}{\lambda^2} - \frac{2a_3}{3\lambda}] {\cal F}_\alpha 
{\cal R}^{\alpha\beta} \zeta_\beta \nonumber \\
 && + [\frac{a_1}{\lambda} + \frac{b_4}{2\lambda^2} - \lambda g_1] 
{\cal C}_{\alpha(2)\beta}  e^\beta{}_{\dot\alpha} {\cal R}^{\alpha(2)}
\zeta^{\dot\alpha} + [g_1 - \frac{b_1}{2\lambda}] 
{\cal F}_{\alpha(2)\beta} {\cal R}^{\alpha(2)} \zeta^\beta.  
\end{eqnarray}
{\bf Spin 2 gauge transformations} In this case we obtain
\begin{eqnarray}
\delta \hat{\cal F}^{\alpha(2)\dot\alpha} &=& - a_1 \eta^{\alpha(2)}
\tilde{\cal F}^{\dot\alpha} - a_2 \xi^{\alpha\dot\alpha}
\tilde{\cal F}^\alpha, \qquad
\delta \hat{\cal F}^\alpha = - a_3 \eta^{\alpha\beta}
\tilde{\cal F}_\beta - a_4\xi^{\alpha\dot\alpha} 
\tilde{\cal F}_{\dot\alpha}, \nonumber \\
\delta \hat{\tilde{\cal F}}^\alpha &=& c_1 {\cal F}^{\alpha\beta(2)}
\eta_{\beta(2)} + c_2 {\cal F}^{\alpha\dot\alpha(2)}
\eta_{\dot\alpha(2)} + c_3 {\cal F}^{\alpha\beta\dot\alpha}
\xi_{\beta\dot\alpha} \\
 && + c_4 \eta^{\alpha\beta} {\cal F}_\beta + c_5
\xi^{\alpha\dot\alpha} {\cal F}_{\dot\alpha} + c_6 
{\cal C}^{\alpha\beta\gamma} e_\beta{}^{\dot\alpha}
\xi_{\gamma\dot\alpha} - c_7 e^\alpha{}_{\dot\alpha}
{\cal C}^{\dot\alpha\dot\beta(2)} \eta_{\dot\beta(2)}. \nonumber
\end{eqnarray}
This gives
\begin{eqnarray}
\delta {\cal L} &=& [\frac{2a_3}{3\lambda} - \frac{c_4}{2\lambda}]
{\cal F}_\alpha \eta^{\alpha\beta} \tilde{\cal F}_\beta
+ [\frac{c_1}{2\lambda} + g_3] {\cal F}_{\alpha(2)\beta}
\eta^{\alpha(2)} \tilde{\cal F}^\beta \nonumber \\
 && + [ - \frac{2a_2}{\lambda} + \frac{c_3}{2\lambda}] 
{\cal F}_{\alpha\beta\dot\alpha} \xi^{\alpha\dot\alpha}
\tilde{\cal F}^\beta - 
[ \frac{2a_2}{\lambda} + \frac{c_6}{\lambda} + 2\lambda^2 g_3]
{\cal C}_{\alpha\beta\gamma} e^\alpha{}_{\dot\alpha}
\xi^{\beta\dot\alpha} \tilde{\cal F}^\gamma. 
\end{eqnarray}
{\bf Hypertransformations} At last we consider
\begin{eqnarray}
\delta \hat{\cal R}^{\alpha(2)} &=& - b_1 \rho^{\alpha(2)\beta}
\tilde{\cal F}_\beta  - b_2 \rho^{\alpha(2)\dot\alpha} 
\tilde{\cal F}_{\dot\alpha}  - b_3 \rho^\alpha \tilde{\cal F}^\alpha,
\nonumber \\
\delta \hat{\tilde{\cal F}}^\alpha &=& - c_1 \rho^{\alpha\beta(2)}
{\cal R}_{\beta(2)}  - c_2 \rho^{\alpha\dot\alpha(2)}
{\cal R}_{\dot\alpha(2)}   c_4 {\cal R}^{\alpha\beta} \rho_\beta
\end{eqnarray}
and obtain
\begin{equation}
\delta {\cal L} = 
[\frac{b_1}{2\lambda^2} - \frac{c_1}{2\lambda} - g_2 ]
\rho_{\alpha(2)\beta} {\cal R}^{\alpha(2)} \tilde{\cal F}^\beta 
+ [- \frac{b_3}{\lambda^2} + \frac{c_4}{2\lambda}] \rho_\alpha 
{\cal R}^{\alpha\beta} \tilde{\cal F}_\beta.
\end{equation}
We have explicitly checked that all coefficients are invariant under
the shifts $\kappa_{1,2}$. The general solution looks like:
\begin{equation}
a_1 = \frac{b_2}{2\lambda} = \frac{c_2}{2}, \qquad
g_1 = \frac{b_1}{2\lambda^2}, \qquad
g_2 = \frac{b_1}{2\lambda^2} - \frac{c_1}{2\lambda}, \qquad
g_3 = - \frac{c_1}{2\lambda}.
\end{equation}

Now we use the fact that field redefinitions do not change the
on-shell part of the cubic vertex. Namely, to calculate a cubic vertex
we choose the solution with $b_1 = c_1 = 0$ where the abelian vertices
are absent. This leaves us with the contribution of the deformed
Lagrangian:
\begin{eqnarray}
{\cal L}_n &=& \frac{1}{\lambda} {\cal F}_{\alpha(2)\dot\alpha}
[ a_1 \Omega^{\alpha(2} \Psi^{\dot\alpha} + a_2 H^{\alpha\dot\alpha}
\Psi^\alpha ]  
 - \frac{2}{3\lambda} {\cal F}_\alpha [ a_3 \Omega^{\alpha\beta}
\Psi_\beta + a_4 H^{\alpha\dot\alpha} \Psi_{\dot\alpha} ] \nonumber \\
 && + \frac{1}{\lambda} {\cal C}_{\alpha(2)\beta} 
e^\beta{}_{\dot\alpha} [ a_1 \Omega^{\alpha(2)} \Psi^{\dot\alpha}
+ a_2 H^{\alpha\dot\alpha} \Psi^\alpha ] \nonumber   \\
 && + \frac{1}{2\lambda^2}  [ - b_2 \Phi_{\alpha(2)\dot\alpha} 
{\cal R}^{\alpha(2)} \Psi^{\dot\alpha} + b_3 \Phi_\alpha 
{\cal R}^{\alpha(2)} \Psi_\alpha + b_4 e^\beta{}_{\dot\alpha}
\phi_{\alpha(2)\beta} {\cal R}^{\alpha(2)} \Psi^{\dot\alpha}]
\nonumber  \\
 && - \frac{1}{2\lambda} [ - c_2 \Phi_{\alpha\dot\alpha(2)}
\Omega^{\dot\alpha(2)} - c_3 \Phi_{\alpha\beta\dot\alpha}
H^{\beta\dot\alpha} + c_4 \Phi^\beta \Omega_{\alpha\beta} + c_5
\Phi^{\dot\alpha} H_{\alpha\dot\alpha} \nonumber \\
 && \qquad + c_6 \phi_{\alpha\beta\gamma} e^\beta{}_{\dot\alpha}
H^{\gamma\dot\alpha} + c_7 e_\alpha{}^{\dot\alpha}
\phi_{\dot\alpha\dot\beta(2)} \Omega^{\dot\beta(2)}] 
\tilde{\cal F}^\alpha. 
\end{eqnarray}
An analysis shows that all the terms with the physical fields are the
same as in the previous subsection, all terms with extra one-form 
$\Phi^{\alpha(3)}$ cancel, so that the only new terms contain
Stueckelberg zero-form $\phi^{\alpha(3)}$:
\begin{eqnarray}
\Delta {\cal L}_1 &=& \frac{4a_0}{3\lambda} \phi_{\alpha\beta(2)}
E^{\beta(2)} [ a_3 \Omega^{\alpha\gamma} \Psi_\gamma + a_4
H^{\alpha\dot\alpha} \Psi_{\dot\alpha} ] \nonumber \\
 && + \frac{1}{\lambda} D \phi_{\alpha(2)\beta} e^\beta{}_{\dot\alpha}
[ a_1 \Omega^{\alpha(2)} \Psi^{\dot\alpha} + a_2 H^{\alpha\dot\alpha}
\Psi^\alpha ] \nonumber \\
 && + \frac{b_4}{2\lambda^2} \phi_{\alpha(2)\beta}
[ D \Omega^{\alpha(2)} + \lambda^2 e^\alpha{}_{\dot\alpha}
H^{\alpha\dot\alpha}] e^\beta{}_{\dot\beta} \Psi^{\dot\beta} \nonumber
\\
 && - \frac{c_6}{2\lambda} \phi_{\alpha\beta\gamma} 
e^\alpha{}_{\dot\alpha} H^{\beta\dot\alpha} \tilde{\cal F}^\gamma 
 + \frac{c_7}{2\lambda} \phi_{\alpha(2)\beta} e^\beta{}_{\dot\alpha}
\Omega^{\alpha(2)} \tilde{\cal F}^{\dot\alpha}.
\end{eqnarray}
This contribution (again up to total derivatives and terms that vanish
on-shell) can be reduces to a very simple form:
\begin{equation}
\Delta {\cal L}_1 = - \frac{2a_1}{\lambda} \phi_{\alpha(2)\beta}
e^\beta{}_{\dot\alpha} \Omega^{\alpha(2)} \tilde{\cal F}^{\dot\alpha}
 - \lambda a_1 \phi_{\alpha(2)\beta} E^{\alpha(2)} H^{\beta\dot\beta}
\Psi_{\dot\beta}. 
\end{equation}
Note that these terms are exactly what we need to compensate the
non-invariance (\ref{noninv}) of the vertex in the previous
subsection.

\section{Partially massless spin 2}

In this case we restrict ourselves with the Skvortsov-Vasiliev
formalism for partially massless spin 5/2. As in all previous cases,
we begin with consistent deformations for all curvatures.\\
{\bf Partially massless spin 5/2} The most general ansatz has the
form:
\begin{eqnarray}
\Delta {\cal F}^{\alpha(2)\dot\alpha} &=& a_1 \Omega^{\alpha(2)}
\Psi^{\dot\alpha} + a_2 H^{\alpha\dot\alpha} \Psi^\alpha  + a_3
B^{\alpha(2)} e_\beta{}^{\dot\alpha} \Psi^\beta + a_4
e^{\alpha\dot\alpha} B^{\alpha\beta} \Psi_\beta + a_5 
e^\alpha{}_{\dot\beta} B^{\dot\alpha\dot\beta} \Psi^\alpha, 
\nonumber \\
\Delta {\cal F}^\alpha &=& a_6 \Omega^{\alpha\beta} \Psi_\beta + a_7
H^{\alpha\dot\alpha} \Psi_{\dot\alpha} + a_8 A \Psi^\alpha + a_9
B^{\alpha\beta} e_{\beta\dot\alpha} \Psi^{\dot\alpha} + a_{10}
e^\alpha{}_{\dot\alpha} B^{\dot\alpha\dot\beta} \Psi_{\dot\beta}.
\end{eqnarray}
General solution for consistency requirement contains three arbitrary
parameters (we take $a_{1,2,6}$):
$$
ma_3 =\lambda a_1 + 2a_2, \qquad
ma_4 = a_2 - \frac{a_0}{3}a_6, \qquad
ma_5 = - \frac{\lambda}{2}a_1 + a_2,
$$
$$
a_7 = - \frac{3\lambda}{2a_0}a_2, \quad
a_8 = \frac{3m}{a_0}a_2, \quad
ma_9 = - a_0a_1 + \frac{3\lambda}{2a_0}a_2 + \lambda a_6, \quad
ma_{10} = - \frac{3\lambda}{2a_0}a_2 - \frac{3\lambda}{2}a_6.
$$
But to understand how many non-trivial parameters we have let us
consider possible field redefinitions:
\begin{eqnarray}
\Phi^{\alpha(2)\dot\alpha} &\Rightarrow& \Phi^{\alpha(2)\dot\alpha}
+ \kappa_1 B^{\alpha(2)} \Psi^{\dot\alpha}, \nonumber \\
\Phi^\alpha &\Rightarrow& \Phi^\alpha + \kappa_2 B^{\alpha\beta}
\Psi_\beta. 
\end{eqnarray}
They make shifts of the deformation parameters:
$$
\Delta a_1 = - m\kappa_1, \qquad
\Delta a_3 = - \lambda \kappa_1, \qquad
\Delta a_5 = \frac{M}{2}\kappa_1, \qquad 
\Delta a_9 = a_0\kappa_1,
$$
$$
\Delta a_4 = \frac{a_0}{3}\kappa_2, \qquad
\Delta a_6 = - m\kappa_2, \qquad
\Delta a_9 = - \lambda \kappa_2, \qquad
\Delta a_{10} = \frac{3M}{2}\kappa_2
$$
and generate abelian vertices
\begin{equation}
\Delta {\cal L}_a = - \frac{\kappa_1}{M} 
{\cal F}_{\alpha(2)\dot\alpha} [ {\cal B}^{\alpha(2)}
\Psi^{\dot\alpha} + B^{\alpha(2)} \tilde{\cal F}^{\dot\alpha}] +
\frac{2\kappa_2}{3M} {\cal F}_\alpha [ {\cal B}^{\alpha\beta}
\Psi_\beta + B^{\alpha\beta} \tilde{\cal F}_\beta]. 
\end{equation}
It is easy to check that general solution is invariant under these
shifts. Therefore, we have only one non-trivial parameter and we use
$a_2$ in what follows.
Special solution obtained in Appendix A.2 corresponds to
\begin{equation}
a_1 = \frac{2}{\lambda}a_2, \qquad
a_6 = - \frac{a_2}{a_0}.
\end{equation}
{\bf Partially massless spin 2} In this case we introduce
\begin{eqnarray}
\Delta {\cal R}^{\alpha(2)} &=& 0, \qquad
\Delta {\cal B}^{\alpha(2)} = 0, \nonumber \\
\Delta {\cal T}^{\alpha\dot\alpha} &=& b_3
\Phi^{\alpha\beta\dot\alpha} \Psi_\beta + b_4 \Phi^{\dot\alpha}
\Psi^\alpha + h.c., \\
\Delta {\cal A} &=& b_6 (\Phi^\alpha \Psi_\alpha + \Phi^{\dot\alpha}
\Psi_{\dot\alpha}) \nonumber
\nonumber
\end{eqnarray}
and obtain
\begin{equation}
b_4 = - \frac{\lambda}{2a_0}b_3, \qquad
b_6 = \frac{m}{a_0}b_3.
\end{equation}
{\bf Massless spin 3/2} Here the most general ansatz looks like:
\begin{eqnarray}
\Delta \tilde{\cal F}^\alpha &=& c_1 \Phi^{\alpha\dot\alpha(2)}
\Omega_{\dot\alpha(2)} + c_2 \Phi^{\alpha\beta\dot\alpha}
H_{\beta\dot\alpha} + c_3 \Phi_\beta \Omega^{\alpha\beta} + c_4
\Phi_{\dot\alpha} H^{\alpha\dot\alpha} + c_5 \Phi^\alpha A 
 + c_6 e^\alpha{}_{\dot\alpha} \Phi^{\beta(2)\dot\alpha}
B_{\beta(2)} \nonumber \\
 && + c_7 e^{\gamma\dot\alpha} \Phi_{\beta\gamma\dot\alpha}
B^{\alpha\beta} + c_8 e_\beta{}^{\dot\alpha} 
\Phi^{\alpha\beta\dot\beta} B_{\dot\alpha\dot\beta} + c_9 
e_\beta{}^{\dot\alpha} \Phi_{\dot\alpha} B^{\alpha\beta} + c_{10}
e^{\alpha\dot\alpha} \Phi^{\dot\beta} B_{\dot\alpha\dot\beta}.
\end{eqnarray}
Here also the general solution has three arbitrary parameters (we take
 $c_{1,2,3}$):
$$
c_4 = - \frac{\lambda}{2a_0}c_2, \qquad 
c_5 = - \frac{m}{a_0}c_2, \qquad 
mc_6 = \lambda c_1 + c_2, \qquad
mc_7 = c_2 - a_0c_3,
$$
$$
mc_8 = - \lambda c_1 + c_2, \qquad
mc_9 = \frac{\lambda}{2a_0}c_2 + \frac{3\lambda}{2}c_3, \qquad
mc_{10} = \frac{2a_0}{3}c_1 + \frac{\lambda}{2a_0}c_2 - \lambda c_3. 
$$
Taking into account possible field redefinitions
\begin{equation}
\Psi^\alpha \Rightarrow \Psi^\alpha + \kappa_3
\Phi^{\alpha\dot\alpha(2)} B_{\dot\alpha(2)} + \kappa_4 \Phi_\beta
B^{\alpha\beta}
\end{equation}
that make shifts of the deformation parameters
$$
\Delta c_1 = m\kappa_3, \qquad
\Delta c_6 = \lambda\kappa_3, \qquad
\Delta c_8 = - M\kappa_3, \qquad
\Delta c_{10} = - \frac{2a_0}{3} \kappa_3,
$$
$$
\Delta c_3 = m\kappa_4, \qquad
\Delta c_7 = - a_0\kappa_4, \qquad
\Delta c_9 = \frac{3M}{2}\kappa_4, \qquad
\Delta c_{10} = \lambda \kappa_4
$$
and generate abelian vertices
\begin{equation}
\Delta {\cal L}_a = - \frac{\kappa_3}{2\lambda} 
[{\cal F}_{\alpha(2)\dot\alpha} B^{\alpha(2)} - 
\Phi_{\alpha(2)\dot\alpha} {\cal B}^{\alpha(2)}] 
\tilde{\cal F}^{\dot\alpha} - \frac{\kappa_4}{2\lambda}
[{\cal F}_\alpha B^{\alpha\beta} - \Phi_\alpha 
{\cal B}^{\alpha\beta}] \tilde{\cal F}_\beta,
\end{equation}
we understand that we have just one non-trivial parameter and we use
$c_2$ in what follows. 

Now let us consider a sum of the deformed free Lagrangian and abelian
vertices
$$
{\cal L} = \hat{\cal L}_0 + {\cal L}_a,
$$
where
\begin{eqnarray}
{\cal L}_a &=& g_1 {\cal F}_{\alpha(2)\dot\alpha} {\cal B}^{\alpha(2)}
\Psi^{\dot\alpha} + g_2 {\cal F}_\alpha {\cal B}^{\alpha\beta}
\Psi_\beta \nonumber \\
 && + g_3 {\cal F}_{\alpha(2)\dot\alpha} B^{\alpha(2)}
\tilde{\cal F}^{\dot\alpha} + g_4 {\cal F}_\alpha B^{\alpha\beta}
\tilde{\cal F}_\beta \nonumber \\
 && + g_5 \Phi_{\alpha(2)\dot\alpha} {\cal B}^{\alpha(2)}
\tilde{\cal F}^{\dot\alpha} + g_6 \Phi_\alpha {\cal B}^{\alpha\beta}
\tilde{\cal F}_\beta, 
\end{eqnarray}
and calculate all its variations. \\
{\bf Supertransformations} Here non-vanishing on-shell variations have
the form:
\begin{eqnarray}
\delta \hat{\cal F}^{\alpha(2)\dot\alpha} &=& a_1 {\cal R}^{\alpha(2)}
\zeta^{\dot\alpha} + a_3 {\cal B}^{\alpha(2)} e_\beta{}^{\dot\alpha}
\zeta^\beta - a_4 e^{\alpha\dot\alpha} {\cal B}^{\alpha\beta}
\zeta_\beta - a_5 e^\alpha{}_{\dot\beta} 
{\cal B}^{\dot\alpha\dot\beta} \zeta^\alpha, \nonumber \\
\delta \hat{\cal F}^\alpha &=& a_6 {\cal R}^{\alpha\beta} \zeta_\beta
+ a_9 {\cal B}^{\alpha\beta} e_{\beta\dot\alpha} \zeta^{\dot\alpha}
- a_{10} e^\alpha{}_{\dot\alpha} {\cal B}^{\dot\alpha\dot\beta} 
\zeta_{\dot\beta}, \\
\delta \hat{\cal T}^{\alpha\dot\alpha} &=& b_3 
{\cal F}^{\alpha\beta\dot\alpha} \zeta_\beta + b_4 
{\cal F}^{\dot\alpha} \zeta^\alpha + h.c. \nonumber
\end{eqnarray}
We obtain
\begin{eqnarray}
\delta {\cal L} &=& [mg_1 - \frac{a_1}{\lambda}] 
{\cal F}_{\alpha(2)\dot\alpha} {\cal R}^{\alpha(2)} \zeta^{\dot\alpha}
+ [\frac{2a_6}{3\lambda} + mg_2] {\cal F}_\alpha 
{\cal R}^{\alpha\beta} \zeta_\beta \nonumber \\
 && + {\cal F}_{\alpha\beta\dot\alpha} [ 
(\frac{a_3}{\lambda} + \frac{b_3}{m} - \lambda g_1) 
e_\gamma{}^{\dot\alpha} {\cal B}^{\alpha\beta} \zeta^\gamma  +
(\frac{2a_4}{\lambda} + \frac{b_3}{m} - a_0g_2)
e^{\beta\dot\alpha} {\cal B}^{\alpha\gamma} \zeta_\gamma ] \nonumber
 \\
 && + [\frac{2a_9}{3\lambda} - \frac{b_2}{m} + \lambda g_2 +
\frac{2a_0}{3}g_1]  {\cal F}_\alpha e_\beta{}^{\dot\alpha} 
{\cal B}^{\alpha\beta} \zeta_{\dot\alpha}. 
\end{eqnarray}
{\bf Spin 2 gauge transformations} Here the variations look like:
\begin{eqnarray}
\delta \hat{\cal F}^{\alpha(2)\dot\alpha} &=& - a_1 \eta^{\alpha(2)}
\tilde{\cal F}^{\dot\alpha} - a_2 \xi^{\alpha\dot\alpha}
\tilde{\cal F}^\alpha, \nonumber \\
\delta \hat{\cal F}^\alpha &=& - a_6 \eta^{\alpha\beta} 
\tilde{\cal F}_\beta - a_7 \xi^{\alpha\dot\alpha} 
\tilde{\cal F}_{\dot\alpha} - a8 \tilde{\cal F}^\alpha \xi, \\
\delta \hat{\tilde{\cal F}}^\alpha &=& c_1 
{\cal F}^{\alpha\dot\alpha(2)} \eta_{\dot\alpha(2)} + c_3 
\eta^{\alpha\beta} {\cal F}_\beta + c_2 
{\cal F}^{\alpha\beta\dot\alpha} \xi_{\beta\dot\alpha} + c_4
\xi^{\alpha\dot\alpha} \tilde{\cal F}_{\dot\alpha} + c_5
{\cal F}^\alpha \xi \nonumber
\end{eqnarray}
and produce
\begin{eqnarray}
\delta {\cal L} &=& 
[\frac{a_1}{\lambda} - \frac{c_1}{2\lambda} - mg_3]
{\cal F}_{\alpha(2)\dot\alpha} \eta^{\alpha(2)} 
\tilde{\cal F}^{\dot\alpha} - 
[\frac{2a_6}{3\lambda} + \frac{c_3}{2\lambda} + mg_4] 
{\cal F}_\alpha \eta^{\alpha\beta} \tilde{\cal F}_\beta \nonumber \\
 && + [\frac{2a_2}{\lambda} + \frac{c_2}{2\lambda}] 
{\cal F}_{\alpha\beta\dot\alpha} \xi^{\alpha\dot\alpha}
\tilde{\cal F}^\beta + [\frac{c_5}{2\lambda} - \frac{2a_8}{3\lambda}]
{\cal F}_\alpha \tilde{\cal F}^\alpha \xi.
\end{eqnarray}
{\bf Hypertransformations} At last we have
\begin{eqnarray}
\delta \hat{\cal T}^{\alpha\dot\alpha} &=& - b_3
\rho^{\alpha\beta\dot\alpha} \tilde{\cal F}_\beta - b_4 \rho^\alpha
\tilde{\cal F}^{\dot\alpha} + h.c., \nonumber \\
\delta \hat{\tilde{\cal F}}^\alpha &=& - c_1
\rho^{\alpha\dot\alpha(2)} {\cal R}_{\dot\alpha(2)} - c_6 
e^\alpha{}_{\dot\alpha} \rho^{\beta(2)\dot\alpha}
{\cal B}_{\beta(2)} - c_7 e^{\gamma\dot\alpha}
\rho_{\beta\gamma\dot\alpha} {\cal B}^{\alpha\beta} - c_8
e_\beta{}^{\dot\alpha} \rho^{\alpha\beta\dot\beta}
{\cal B}_{\dot\alpha\dot\beta} \\
 && - c_3 \rho_\beta {\cal R}^{\alpha\beta} - c_9 \rho_{\dot\alpha}
e_\beta{}^{\dot\alpha} {\cal B}^{\alpha\beta} - c_{10}
\rho^{\dot\alpha} e^{\alpha\dot\beta} {\cal B}_{\dot\alpha\dot\beta}
\nonumber
\end{eqnarray}
and obtain
\begin{eqnarray}
\delta {\cal L} &=&  [\frac{c_1}{2\lambda} - mg_5] 
\rho_{\alpha(2)\dot\alpha} {\cal R}^{\alpha(2)} 
\tilde{\cal F}^{\dot\alpha} + [\frac{c_3}{2\lambda} - mg_6]
\rho_\alpha {\cal R}^{\alpha\beta} \tilde{\cal F}_\beta \nonumber \\
 && + \rho_{\alpha\beta\dot\alpha} 
[ (\frac{c_6}{2\lambda} - \frac{b_3}{m})
e_\gamma{}^{\dot\alpha} {\cal B}^{\alpha\beta} \tilde{\cal F}^\gamma
+ (\frac{c_7}{2\lambda} - \frac{b_3}{m} + a_0g_6) e^{\beta\dot\alpha} 
{\cal B}^{\alpha\gamma}  \tilde{\cal F}_\gamma ]\nonumber \\
 && + [\frac{3\lambda}{2}g_6 - \frac{b_4}{m} - \frac{c_9}{2\lambda}]
\rho_\alpha e^\alpha{}_{\dot\alpha} {\cal B}^{\dot\alpha\dot\beta}
\tilde{\cal F}_{\dot\beta}. 
\end{eqnarray}
We have explicitly checked that all the coefficients are invariant
under the shifts $\kappa_{1,2,3,4}$. The general solution is:
$$
g_1 = \frac{a_1}{m\lambda}, \qquad
g_2 = - \frac{2a_6}{3\lambda}, \qquad
g_3 = \frac{a_1}{m\lambda} - \frac{c_1}{2m\lambda},
$$
$$
g_4 = - \frac{2a_6}{3m\lambda} - \frac{c_3}{2m\lambda}, \qquad
g_5 = \frac{c_1}{2m\lambda}, \qquad
g_6 = \frac{c_3}{2m\lambda},
$$
$$
a_2 = - \frac{\lambda}{2}b_3, \qquad
c_2 = 2\lambda b_3.
$$

To calculate cubic vertex we again use the freedom related with field
redefinitions and take the solution at $a_1 = a_6 = c_1 = c_3 = 0$ so
that there are no abelian vertices:
\begin{eqnarray}
{\cal L}_1 &=& - \frac{1}{\lambda} {\cal F}_{\alpha(2)\dot\alpha}
[ a_2 H^{\alpha\dot\alpha} \Psi^\alpha + a_3 B^{\alpha(2)}
e_\beta{}^{\dot\alpha} \Psi^\beta + a_4 e^{\alpha\dot\alpha}
B^{\alpha\beta} \Psi_\beta ] \nonumber \\
 && + \frac{2}{3\lambda} {\cal F}_\alpha [ a_7 H^{\alpha\dot\alpha}
\Psi_{\dot\alpha} + a_8 A \Psi^\alpha + a_9 B^{\alpha\beta}
e_{\beta\dot\alpha} \Psi^{\dot\alpha} ] \nonumber \\
 && + \frac{1}{m} [ {\cal B}^{\alpha\beta} e_\beta{}^{\dot\alpha}-
{\cal B}^{\dot\alpha\dot\beta} e^\alpha{}_{\dot\beta}]
[ - b_1 \Phi_{\alpha\gamma\dot\alpha} \Psi^\gamma + b_2
\Phi_{\dot\alpha} \Psi_\alpha ] \nonumber \\
 && - \frac{1}{2\lambda} [ c_2 \Phi^{\alpha\beta\dot\alpha}
H_{\beta\dot\alpha} + c_4 \Phi_{\dot\alpha}H^{\alpha\dot\alpha}
+ c_5 \Phi^\alpha A \nonumber \\
 && \qquad + c_7 e^{\gamma\dot\alpha} \Phi_{\beta\gamma\dot\alpha}
B^{\alpha\beta} + c_8 e_\beta{}^{\dot\alpha} 
\Phi^{\alpha\beta\dot\beta} B_{\dot\alpha\dot\beta} + c_9
e_\beta{}^{\dot\alpha} \Phi_{\dot\alpha} B^{\alpha\beta}] 
\tilde{\cal F}_\alpha. 
\end{eqnarray}
To simplify a comparison with the results of previous Section we
present our final result in the unitary gauge $B^{\alpha(2)} = 0$:
\begin{eqnarray}
{\cal L}_1 &=& \frac{a_2}{a_0} \Phi_\alpha H^{\alpha\dot\alpha} 
\tilde{\cal F}_{\dot\alpha} - \frac{5m}{a_0}a_2 \Phi_{\dot\alpha}
e^{\alpha\dot\alpha} A \Psi_\alpha \nonumber \\
 && + \frac{4a_2}{\lambda} \Phi_{\alpha\beta\dot\alpha} 
e^\alpha{}_{\dot\beta} \Omega^{\dot\alpha\dot\beta} \Psi^\beta
 + 2a_2 e_{(\alpha}{}^{\dot\alpha} \Phi_{\beta)\dot\alpha\dot\beta}
H^{\alpha\dot\beta} \Psi^\beta + a_2 e^{\alpha\dot\alpha}
\Phi_{\alpha\dot\alpha\dot\beta} H^{\beta\dot\beta} \Psi_\beta
\nonumber \\
 && - \frac{a_2}{a_0} \Phi_{\dot\alpha} (e_\beta{}^{\dot\alpha}
\Omega^{\alpha\beta} + \Omega^{\dot\alpha\dot\beta} 
e^\alpha{}_{\dot\beta}) \Psi_\alpha - \frac{a_0}{3\lambda}a_2
\Phi_\alpha e^{(\alpha}{}_{\dot\alpha} H^{\beta)\dot\beta} \Psi_\beta
 - \frac{a_0}{15\lambda}a_2 \Phi_\alpha e_{\beta\dot\alpha}
H^{\beta\dot\alpha} \Psi^\alpha.
\end{eqnarray}

\section{Final remarks}

In this work we considered a partially massless spin 5/2
supermultiplet which contains partially massless spin 5/2, massless
and partially massless spin 2, as well as massless spin 3/2. As with
any supermultiplet, the important question is whether it can interact
with supergravity. As a first step, we considered global
supertransformations that connect partially massless spin 5/2 with its
two possible superpartners, massless and partially massless spin 2. We
then made these transformations local by coupling to massless
gravitino.

We use a frame-like gauge-invariant multispinor formalism to describe
free fields. Starting with spin 5/2, this formalism requires the
introduction of so-called extra fields which do not appear in the free
Lagrangian, but are necessary to construct a complete set of gauge
invariant objects. As a result, the usual constructive approach to
interactions can't be used, and we apply the so-called 
Fradkin-Vasiliev formalism. There are two possible descriptions
of partially massless spin 5/2: a Skvortsov-Vasiliev one and what we
call general formalism, which comes from a massive case at a special
value of mass. For the case with massless spin 2 as a superpartner, we
demonstrated that these two approaches produce completely consistent
results.

As it is common for gauge-invariant formalism for massive and
partially massless fields, we encounter ambiguities associated with
the
field redefinitions involving Stueckelberg fields. In the
Fradkin-Vasiliev formalism, they generate combinations of abelian and 
non-abelian vertices that vanish on-shell and don't change the
on-shell part of a vertex. We use this freedom to simplify
calculations. At the same time, we demonstrated that ambiguities in
supertransformations can be resolved using unfolded equations.
However, how to fix ambiguities for the gravitino remains an open
question.

\appendix

\section{Supersymmetry}

In this Appendix we consider a consistent deformation for the unfolded
equations with massless or partially massless spin 2 as a superpartner
of the partially massless spin 5/2. Recall that a complete set of
the unfolded equations contains three sectors: one-forms, Stueckelberg
zero-forms and infinite set of the gauge invariant zero-forms.The last
sector forms a closed subsystem and it appears that the deformation
procedure gives a unique result without any ambiguities related with
field redefinitions. Then this result can be consistently extended to
two other sectors.

\subsection{Massless spin 2}

In this subsection we consider massless spin 2 as a superpartner.\\
{\bf Partially massless spin 5/2} We  begin with the sector of the
gauge invariant zero-forms. The most general ansatz for their unfolded
equations looks like:
\begin{eqnarray}
0 &=& D Y^{\alpha(5+k)\dot\alpha(k)} + e_{\beta\dot\beta}
Y^{\alpha(5+k)\beta\dot\alpha(k)\dot\beta} + c_{1,k}
e^\alpha{}_{\dot\beta} \phi^{\alpha(4+k)\dot\alpha(k)\dot\beta}
+ d_{1,k} e^{\alpha\dot\alpha} Y^{\alpha(4+k)\dot\alpha(k-1)}
\nonumber \\
 && + \gamma_{1,k} W^{\alpha(5+k)\dot\alpha(k)\dot\beta}
\Psi_{\dot\beta} + \delta_{1,k} W^{\alpha(4+k)\dot\alpha(k)}
\Psi^\alpha, \\
0 &=& D \phi^{\alpha(4+k)\dot\alpha(k+1)} + e_{\beta\dot\beta}
\phi^{\alpha(4+k)\beta\dot\alpha(k+1)\dot\beta} + c_{2,k}
e_\beta{}^{\dot\alpha} Y^{\alpha(4+k)\beta\dot\alpha(k)} + d_{2,k}
e^{\alpha\dot\alpha} \phi^{\alpha(3+k)\dot\alpha(k)} \nonumber \\
 && + \gamma_{2,k} W^{\alpha(4+k)\beta\dot\alpha(k+1)} \Psi_\beta
+ \delta_{2,k} W^{\alpha(4+k)\dot\alpha(k)} \Psi^{\dot\alpha}. 
\end{eqnarray}
Consistency requires
$$
\gamma_{1,k} = \gamma_1, \qquad \gamma_{2,k} = \gamma_2, \qquad
\delta_{1,k} = \frac{(k+2)}{(k+5)}\delta_1, \qquad
\delta_{2,k} = \frac{(k+6)}{(k+1)}\delta_2,
$$
$$
\gamma_1 = - 5\lambda\gamma_2, \qquad
\delta_1 = - 5\lambda^2\gamma_2, \qquad
\delta_2 = \lambda\gamma_2.
$$
We proceed with the equations for the extra fields because they are
directly connected with the sector of the gauge invariant zero-forms.
The deformed equations have the form:
\begin{eqnarray}
0 &=& D \Phi^{\alpha(3)} + 2\lambda^2 E^\alpha{}_\beta
\phi^{\alpha(2)\beta} + E_{\beta(2)} Y^{\alpha(3)\beta(2)} 
 + g_1 e_{\beta\dot\alpha} W^{\alpha(3)\beta} \Psi^{\dot\alpha},
\nonumber \\
0 &=& D \phi^{\alpha(3)} - \Phi^{\alpha(3)} + e_{\beta\dot\alpha}
\phi^{\alpha(3)\beta\dot\alpha} + g_2 W^{\alpha(3)\beta} \Psi_\beta. 
\end{eqnarray}
We obtain
$$
g_1 = \gamma_1, \qquad g_2 = \gamma_2.
$$
Then we consider deformed equations for the physical fields:
\begin{eqnarray}
0 &=& D \Phi^{\alpha(2)\dot\alpha} + e_\beta{}^{\dot\alpha}
\Phi^{\alpha(2)\beta} + \frac{M}{2} e^\alpha{}_{\dot\beta}
\Phi^{\alpha\dot\alpha\dot\beta} + \frac{a_0}{3} e^{\alpha\dot\alpha}
\Phi^\alpha \nonumber \\
 && + a_1 \Omega^{\alpha(2)} \Psi^{\dot\alpha} + a_2
H^{\alpha\dot\alpha} \Psi^\alpha,   \\
0 &=& D \Phi^\alpha + a_0 e_{\beta\dot\alpha}
\Phi^{\alpha\beta\dot\alpha} + \frac{3M}{2} e^\alpha{}_{\dot\alpha}
\Phi^{\dot\alpha} - 2a_0 E_{\beta(2)} \phi^{\alpha\beta(2)} \nonumber
\\
 && + a_3 \Omega^{\alpha\beta} \Psi_\beta + a_4 H^{\alpha\dot\alpha}
\Psi_{\dot\alpha} 
\end{eqnarray}
and obtain
\begin{equation}
a_1 = - g_1, \quad M = - \lambda, \quad
a_2 = - \frac{\lambda}{2}a_1, \quad 
a_3 = - \frac{3\lambda}{2a_0}a_1, \quad
a_4 = - \frac{3a_0}{5}a_1.
\end{equation}
{\bf Massless spin 2} Here we also begin with the equations for the
gauge invariant zero-forms. The most general  ansatz for their
deformation looks like:
\begin{eqnarray}
0 &=& D W^{\alpha(4+k)\dot\alpha(k)} + e_{\beta\dot\beta}
W^{\alpha(4+k)\beta\dot\alpha(k)\dot\beta} + \lambda^2
e^{\alpha\dot\alpha} W^{\alpha(3+k)\dot\alpha(k-1)} \nonumber \\
 && + \alpha_{1,k} Y^{\alpha(4+k)\beta\dot\alpha(k)} \Psi_\beta +
\beta_{1,k} Y^{\alpha(4+k)\dot\alpha(k-1)} \Psi^{\dot\alpha} \nonumber
\\
 && + \alpha_{2,k} \phi^{\alpha(4+k)\dot\alpha(k)\dot\beta}
\Psi_{\dot\beta} + \beta_{2,k} \phi^{\alpha(3+k)\dot\alpha(k)}
\Psi^\alpha. 
\end{eqnarray}
Consistence gives
$$
\alpha_{1,k} = \alpha_1, \qquad \alpha_{2,k} = \alpha_2, \qquad 
\beta_{1,k} = \frac{(k+4)}{(k+1)}\beta_1, \qquad
\beta_{2,k} = \frac{k}{(k+5)}\beta_2,
$$
$$
\alpha_2 = 3\lambda\alpha_1, \qquad
\beta_1 = \lambda\alpha_1, \qquad
\beta_2 = 3\lambda^2\alpha_1.
$$
We proceed with the equations for  the gauge sector. The most general
ansatz has the form:
\begin{eqnarray}
0 &=& D \Omega^{\alpha(2)} +  \lambda^2 e^\alpha{}_{\dot\alpha}
H^{\alpha\dot\alpha} + E_{\beta(2)} W^{\alpha(2)\beta(2)} \nonumber \\
 && + b_1 \Omega^{\alpha(2)\beta} \Psi_\beta + b_2 
\Phi^{\alpha(2)\dot\alpha} \Psi_{\dot\alpha} + b_3 \Phi^\alpha
\Psi^\alpha + b_4 e_{\beta\dot\alpha} \phi^{\alpha(2)\beta}
\Psi^{\dot\alpha} + b_0 e_{\beta\dot\alpha}
 \phi^{\alpha(2)\beta\gamma\dot\alpha} \Psi_\gamma, \nonumber \\
0 &=& D H^{\alpha\dot\alpha} + e_\beta{}^{\dot\alpha}
\Omega^{\alpha\beta} + e^\alpha{}_{\dot\beta}
\Omega^{\dot\alpha\dot\beta}  \\
 && + b_5 (\Phi^{\alpha\beta\dot\alpha} \Psi_\beta 
+ \Phi^{\alpha\dot\alpha\dot\beta} \Psi_{\dot\beta}) + b_6
(\Phi^\alpha \Psi^{\dot\alpha} + \Phi^{\dot\alpha} \Psi^\alpha) 
+ b_7 (e_\beta{}^{\dot\alpha} \phi^{\alpha\beta\gamma} \Psi_\gamma +
e^\alpha{}_{\dot\beta} \phi^{\dot\alpha\dot\beta\dot\gamma}
\Psi_{\dot\gamma}). \nonumber
\end{eqnarray}
We get
\begin{equation}
b_3 = - \frac{\lambda}{2a_0}b_2, \quad 
b_4 = \frac{3}{2}b_2, \quad
b_1 = b_5 = - \frac{1}{2\lambda}b_2, \quad
b_6 = \frac{3}{4a_0}b_2, \quad
b_0 = b_7 = 0.
\end{equation}

\subsection{Partially massless spin 2}

In this subsection we consider partially massless spin 2 as a
superpartner. \\
{\bf Partially massless spin 5/2} The most general ansatz for the
gauge invariant zero-forms looks like:
\begin{eqnarray}
0 &=& D Y^{\alpha(5+k)\dot\alpha(k)} + e_{\beta\dot\beta}
Y^{\alpha(5+k)\beta\dot\alpha(k)\dot\beta} + c_{1,k}
e^\alpha{}_{\dot\beta} \phi^{\alpha(4+k)\dot\alpha(k)\dot\beta}
+ d_{1,k} e^{\alpha\dot\alpha} Y^{\alpha(4+k)\dot\alpha(k-1)},
\nonumber \\
 && + \gamma_{1,k} W^{\alpha(5+k)\dot\alpha(k)\dot\beta}
\Psi_{\dot\beta} + \delta_{1,k} W^{\alpha(4+k)\dot\alpha(k)}
\Psi^\alpha \\
0 &=& D \phi^{\alpha(4+k)\dot\alpha(k+1)} + e_{\beta\dot\beta}
\phi^{\alpha(4+k)\beta\dot\alpha(k+1)\dot\beta} + c_{2,k}
e_\beta{}^{\dot\alpha} Y^{\alpha(4+k)\beta\dot\alpha(k)} + d_{2,k}
e^{\alpha\dot\alpha} \phi^{\alpha(3+k)\dot\alpha(k)} \nonumber \\
 && + \gamma_{2,k} W^{\alpha(4+k)\beta\dot\alpha(k+1)} \Psi_\beta
+ \delta_{2,k} W^{\alpha(4+k)\dot\alpha(k)} \Psi^{\dot\alpha}
\nonumber \\
 && + \gamma_{3,k} B^{\alpha(4+k)\dot\alpha(k+1)\dot\beta}
\Psi_{\dot\beta} + \delta_{3,k} B^{\alpha(3+k)\dot\alpha(k+1)}
\Psi^\alpha. 
\end{eqnarray}
Consistency requires
$$
\gamma_{1,k} = \gamma_1, \qquad
\gamma_{2,k} = \gamma_2 = - \frac{\gamma_1}{15\lambda}, \qquad
\gamma_{3,k} = \gamma_3 = \frac{4m}{15\lambda^2}\gamma_1,
$$
$$
\delta_{1,k} = \frac{(k+4)}{(k+5)}\lambda\gamma_1, \qquad
\delta_{2,k} = \frac{(k+4)(k+6)}{(k+1)(k+2)}\lambda\gamma_2, \qquad 
\delta_{3,k} = \frac{(k+6)}{(k+5)} \lambda\gamma_3.
$$
We proceed with the equations for the extra fields:
\begin{eqnarray}
0 &=& D \Phi^{\alpha(3)} + 2\lambda^2 E^\alpha{}_\beta 
\phi^{\alpha(2)\beta} + E_{\beta(2)} Y^{\alpha(3)\beta(2)} \nonumber 
\\
 && + mg_0 \Omega^{\alpha(2)} \Psi^\alpha + \lambda g_0
e^\alpha{}_{\dot\alpha} B^{\alpha(2)} \Psi^{\dot\alpha} 
 + f_1 e_{\beta\dot\alpha} W^{\alpha(3)\beta} \Psi^{\dot\alpha}
+ f_2 e_{\beta\dot\alpha} B^{\alpha(2)\beta\dot\alpha} \Psi^\alpha, \\
0 &=& D \phi^{\alpha(3)} - \Phi^{\alpha(3)} + e_{\beta\dot\alpha}
\phi^{\alpha(3)\beta\dot\alpha} \nonumber \\
 && + g_0 B^{\alpha(2)} \Psi^\alpha + g_1 W^{\alpha(3)\beta}
\Psi_\beta + g_2 B^{\alpha(3)\dot\alpha} \Psi_{\dot\alpha}
\end{eqnarray}
and obtain
$$
g_0 = \frac{m}{3\lambda}\gamma_1, \qquad
g_1 = \gamma_2, \qquad g_2 = \gamma_3, \qquad
f_1 = \gamma_1, \qquad f_2 = 0.
$$
Then the deformed equations for the physical fields have the form:
\begin{eqnarray}
0 &=& D \Phi^{\alpha(2)\dot\alpha} + e_\beta{}^{\dot\alpha}
\Phi^{\alpha(2)\beta} + \frac{M}{2} e^\alpha{}_{\dot\beta}
\Phi^{\alpha\dot\alpha\dot\beta} + \frac{a_0}{3} e^{\alpha\dot\alpha}
\Phi^\alpha \nonumber \\
 && + a_1 \Omega^{\alpha(2)} \Psi^{\dot\alpha} + a_2
H^{\alpha\dot\alpha} \Psi^\alpha + a_3 B^{\alpha(2)} 
e_\beta{}^{\dot\alpha} \Psi^\beta + a_4 e^{\alpha\dot\alpha}
B^{\alpha\beta} \Psi_\beta + a_5 e^\alpha{}_{\dot\beta}
B^{\dot\alpha\dot\beta} \Psi^\alpha, \nonumber  \\
0 &=& D \Phi^\alpha + a_0 e_{\beta\dot\alpha}
\Phi^{\alpha\beta\dot\alpha} + \frac{3M}{2} e^\alpha{}_{\dot\alpha}
\Phi^{\dot\alpha} - 2a_0 E_{\beta(2)} \phi^{\alpha\beta(2)} \\
 && + a_6 \Omega^{\alpha\beta} \Psi_\beta + a_7
H^{\alpha\dot\alpha} \Psi_{\dot\alpha} + a_8 A \Psi^\alpha + a_9
B^{\alpha\beta} e_{\beta\dot\alpha} \Psi^{\dot\alpha} + a_{10}
e^\alpha{}_{\dot\alpha} B^{\dot\alpha\dot\beta} \Psi_{\dot\beta}.
\nonumber,
\end{eqnarray}
We obtain
$$
M = \lambda, \qquad a_1 = - \gamma_1, \qquad
a_3 = a_4 = a_5 = a_{10} = 0,
$$
\begin{equation}
a_2 = \frac{\lambda}{2}a_1, \qquad
a_6 = - \frac{\lambda a_1}{2a_0}, \qquad
a_7 = - \frac{3\lambda^2}{4a_0}a_1, \qquad
a_8 = \frac{3m\lambda}{2a_0}a_1, \qquad
a_9 = - \frac{2m}{a_0}a_1.
\end{equation}
{\bf Partially massless spin 2} The most general ansatz for the gauge
invariant zero-forms looks like:
\begin{eqnarray}
0 &=& D W^{\alpha(4+k)\dot\alpha(k)} + e_{\beta\dot\beta}
W^{\alpha(4+k)\beta\dot\alpha(k)\dot\beta} + a_{1,k} 
e^\alpha{}_{\dot\beta} B^{\alpha(3+k)\dot\alpha(k)\dot\beta}
+ b_{1,k} e^{\alpha\dot\alpha} W^{\alpha(3+k)\dot\alpha(k-1)}
\nonumber \\
 && + \alpha_{1,k} Y^{\alpha(4+k)\beta\dot\alpha(k)} \Psi_\beta +
\beta_{1,k} Y^{\alpha(4+k)\dot\alpha(k-1)} \Psi^{\dot\alpha} \nonumber
\\
 && + \alpha_{2,k} \phi^{\alpha(4+k)\dot\alpha(k)\dot\beta}
\Psi_{\dot\beta} + \beta_{2,k} \phi^{\alpha(3+k)\dot\alpha(k)}
\Psi^\alpha,  \\
0 &=& D B^{\alpha(3+k)\dot\alpha(k+1)} + e_{\beta\dot\beta}
B^{\alpha(3+k)\beta \dot\alpha(k+1)\dot\beta} + a_{2,k}
e_\beta{}^{\dot\alpha} W^{\alpha(3+k)\beta\dot\alpha(k)} + 
b_{2,k} e^{\alpha\dot\alpha} B^{\alpha(2+k)\dot\alpha(k)} \nonumber \\
 && + \alpha_{3,k} \phi^{\alpha(3+k)\beta\dot\alpha(k+1)} \Psi_\beta 
+ \beta_{3,k} \phi^{\alpha(3+k)\dot\alpha(k)} \Psi^{\dot\alpha}.
\end{eqnarray}
Consistency requires
$$
\alpha_{1,k} = \alpha_1, \qquad 
\alpha_{2,k} = \alpha_2 = \lambda \alpha_1, \qquad
\alpha_{3,k} = \alpha_3 = - m\alpha_1, 
$$
$$
\beta_{1,k} = \frac{(k+2)}{(k+1)}\lambda\alpha_1, \qquad
\beta_{2,k} = \frac{k(k+2)}{(k+4)(k+5)}\lambda\alpha_2, \qquad 
\beta_{3,k} = \frac{k}{(k+1)}\lambda\alpha_3.
$$
Equations for the auxiliary fields have the form:
\begin{eqnarray}
0 &=& D \Omega^{\alpha(2)} + m E^\alpha{}_\beta B^{\alpha\beta} 
+ E_{\beta(2)} W^{\alpha(2)\beta(2)} \nonumber \\
 && + b_1 \Phi^{\alpha(2)\beta} \Psi_\beta + \lambda b_1
\phi^{\alpha(2)\beta} e_\beta{}^{\dot\alpha} \Psi_{\dot\alpha} +
g_0 e_{\beta\dot\alpha} \phi^{\alpha(2)\beta\gamma\dot\alpha}
\Psi_\gamma, \nonumber \\
0 &=& D B^{\alpha(2)} + m \Omega^{\alpha(2)} + e_{\beta\dot\alpha}
B^{\alpha(2)\beta\dot\alpha} \\
 && + b_7 \phi^{\alpha(2)\beta} \Psi_\beta. \nonumber 
\end{eqnarray}
We  obtain
$$
b_1 = - \alpha_1, \qquad
b_2 = \lambda b_1, \qquad
b_7 = mb_1, \qquad
g_0 = 0.
$$
Equations for the physical fields
\begin{eqnarray}
0 &=& D H^{\alpha\dot\alpha} + e_\beta{}^{\dot\alpha}
\Omega^{\alpha\beta} + e^\alpha{}_{\dot\beta}
\Omega^{\dot\alpha\dot\beta} + m e^{\alpha\dot\alpha} A \nonumber \\
 &&  + b_3 \Phi^{\alpha\beta\dot\alpha} \Psi_\beta + b_4 
\Phi^{\dot\alpha} \Psi^\alpha + b_5 e_\beta{}^{\dot\alpha}
\phi^{\alpha\beta\gamma} \Psi_\gamma + h.c., \nonumber \\
0 &=& D A + 2 E_{\alpha(2)} B^{\alpha(2)} + 2 E_{\dot\alpha(2)}
B^{\dot\alpha(2)} + m e_{\alpha\dot\alpha} H^{\alpha\dot\alpha} \\
 && + b_6 (\Phi^\alpha \Psi_\alpha + \Phi^{\dot\alpha}
\Psi_{\dot\alpha}). \nonumber 
\end{eqnarray}
We get
$$
b_3 = b_1, \qquad
b_4 = - \frac{\lambda}{2a_0}b_1, \qquad
b_5 = 0, \qquad
b_6 = \frac{m}{a_0}b_1.
$$

%\bibliography{biblio}
%\bibliographystyle{zplain}

\end{document}